\documentclass[times,twocolumn,final,longtitle]{elsarticle}

\usepackage{jasr}
\usepackage{framed,multirow}

\usepackage{amssymb}
\usepackage{latexsym}

\usepackage[switch]{lineno}

\usepackage{url}
\usepackage{xcolor}
\usepackage[table]{xcolor}
\definecolor{newcolor}{rgb}{.8,.349,.1}

\usepackage[citebordercolor=blue]{hyperref}

\usepackage{dcolumn}
\newcolumntype{d}{D{.}{.}{3.2}}

\journal{Advances in Space Research}

\begin{document}

\verso{L. Schulz \textit{et al.}}

\begin{frontmatter}

\title{Space waste: An update of the anthropogenic matter injection into Earth atmosphere}%

\author[1]{Leonard \snm{Schulz}\corref{cor1}}
\ead{l.schulz@tu-bs.de}
\cortext[cor1]{Corresponding author.}
\author[1,2]{Karl-Heinz \snm{Glassmeier}}
\ead{kh.glassmeier@tu-bs.de}
\author[3]{Moritz \snm{Herberhold}}
\ead{moritz.herberhold@dlr.de}
\author[4]{Adam \snm{Mitchell}}
\ead{adam.mitchell@ext.esa.int}
\author[5,6]{Daniel M. \snm{Murphy}}
\ead{daniel.m.murphy@noaa.gov}
\author[7]{John M. C. \snm{Plane}}
\ead{j.m.c.plane@leeds.ac.uk}
\author[1]{Ferdinand \snm{Plaschke}}
\ead{f.plaschke@tu-bs.de}

\affiliation[1]{organization={Institute of Geophysics and Extraterrestrial Physics, Technische Universität Braunschweig},
                city={Braunschweig},
                postcode={38106},
                country={Germany}}

\affiliation[2]{organization={Max-Planck-Institute of Solar System Science},
                city={G\"ottingen},
                postcode={37077},
                country={Germany}}
\affiliation[3]{organization={German Aerospace Center (DLR)},
                city={Bremen},
                postcode={28359},
                country={Germany}}
\affiliation[4]{organization={European Space Agency (ESA), European Space Research and Technology Centre},
                city={Noordwijk},
                postcode={2200 AG},
                country={Netherlands}}
\affiliation[5]{organization={Chemical Sciences Laboratory, National Oceanic and Atmospheric Administration},
                city={Boulder, CO},
                postcode={80305},
                country={USA}}
\affiliation[6]{organization={Purdue University},
                city={West Lafayette, IN},
                postcode={47907},
                country={USA}}
\affiliation[7]{organization={University of Leeds},
                city={Leeds},
                postcode={LS2 9JT},
                country={United Kingdom}}


\begin{abstract}
Large satellite constellations are one of the main reasons for an increasing amount of mass being brought into low Earth orbit in recent years. After end of life, the satellites, as well as rocket stages, reenter Earth's atmosphere. This space waste burns up in the atmosphere and thus injects a substantial amount of its matter into the mesosphere and lower thermosphere. A first comprehensive analysis of the anthropogenic injection and a comparison to the natural injection by meteoroids was presented by \citet{art_Schulz_2021} (Advances in Space Research, 67 (3), 1002-1025). They found significant and even the dominant injection of several metal elements regularly used in spacecraft compared to the natural injection. The first observations of space waste remnants in stratospheric aerosol particles \citep[][PNAS, Vol. 120, No. 43, e2313374120]{art_Murphy_2023} confirmed several of these estimates, but also revealed differences and new insights: in particular, a much more extensive range of elements. These inventories and observations raise questions of possible environmental effects, especially on the stratospheric ozone layer \citep[][Understanding the Atmospheric Effects from Spacecraft Re-entry, ESA, Whitepaper, 2024]{techrep_Mitchell_2024}. The current study presents an update to the space waste injection estimates of \citet{art_Schulz_2021}, assessing the years from 2015 to 2025 using available reentry databases but also considering future mass influx scenarios. Combining mass influx data with a detailed estimates of average satellite and rocket stage composition as well as ablation factors allows the estimation of the mass injection of 43 elements and thus a much more detailed comparison to the meteoric injection. Comparison of estimated elemental fluxes to stratospheric aerosol data shows excellent agreement. In general, from 2020 onward, a strong rise in space waste mass influx to the atmosphere can be seen, which has led to more than double the atmospheric injected mass in 2024 compared to levels from 2015 to 2020. Future scenarios discussed by \citet{art_Schulz_2021} may already be reached by the end of 2025. In 2024, 24 elements were dominating the meteoric injection compared to 18 in 2015, this might increase to 30 in the future. Several of them are transition metals, which are known for their catalytic activity. This indicates a substantial risk of long-term adverse effects on the atmosphere such as ozone depletion, radiative effects and changes in cloud formation, if no action is taken. Research is urgently needed into the atmospheric accumulation, chemistry, and general atmospheric effects of specific elements. 
\end{abstract}


\end{frontmatter}


\section{Introduction}
\label{sec1}
The marked increase in space activities in the past 5 years has led to a strong rise in mass that is entering into orbit. The main driver is large satellite constellations with thousands of satellites already launched, and tens- to hundreds of thousands being planned. All in all, more than a million constellation satellites are proposed \citep{art_Falle_2023}. This creates and enhances three different sustainability problems of spaceflight:
\begin{enumerate}
    \item \textit{Space debris} is ``artificial material that is orbiting Earth but no longer functional'' \citep{mics_EncyclopBrit_SpaceDebris}. It is a long-known problem \citep[e.\,g.][]{Book_2006_Klinkrad} posed by satellites and rocket stages staying in orbit after their end of life. The largely increased number of objects, especially in orbital shells occupied by large satellite constellation, increases the risk of collisions. In the worst case, this leads to the Kessler syndrome, in which a long-term chain reaction of collisions can render entire orbital shells unusable. This issue is being addressed by space agencies, which project worsening conditions even if no further launches were to take place \citep{techrep_ESA_SpaceDebris_2025}. For LEO (Low Earth Orbit), the only viable mitigation option is the reentry of an object into the atmosphere. This leads to another problem, namely 
    \item  \textit{Ground impacts}. Parts of satellites and rocket bodies can survive destructive reentry \citep[e.\,g.][]{inPro_Ailor_2005} and end up on the ground. As the number of reentering objects rises, the total risk of a deadly event increases as well \citep{art_Pardini_2025}. To minimize this risk, efforts are being made toward complete demise of spacecraft upon atmospheric reentry. This risk minimization amplifies the third problem, namely
    \item \textbf{Space waste}, which we define as all orbital and suborbital man-made objects (re)entering Earth's atmosphere and destructively ablating. This way, considerable amounts of vapor and particulates are released into the upper atmosphere. We use the term space waste to contrast with the well-known problem of space debris, which focuses on effects in orbit. Space waste on the other hand has effects on Earth's atmosphere. 
\end{enumerate}
The atmospheric influence from space waste reentry has long been regarded as negligible \citep{tech_Lohn_1994, tech_Smith_1999, techrep_ATISPADE_2019, techrep_ARA_2019}. However, \citet{art_Schulz_2021} showed that the mass annually injected to the atmosphere is already significant compared to the natural input by ablation of meteoroids and can even dominate for some metals like Al or Cu. Measurements of space waste remnants in stratospheric aerosol particles \citep{art_Murphy_2023} largely confirmed these estimates and also showed Li and Pb dominating. A significance or dominance compared to natural levels inevitably poses the question of possible environmental effects. This is especially true for Earth's atmosphere due to the complexity of the system: the same constituent can have different effects in different atmospheric layers. Space waste can start ablating at altitudes as high as 110\,km and ablation may last down to 40\,km \citep[][and references therein]{art_Yamauchi_2024}. However, the majority of the material is injected into the mesosphere, where it is expected to mix with the meteoroid background to form aerosol particles. Mesospheric (50 to 85\,km) circulation transports these aerosols to the winter poles, where they descend toward the stratosphere and can be measured \citep{art_Murphy_2023}. The wide variety of space waste metals that have low abundances in meteoroids means that new chemistry is expected in the mesosphere and stratosphere. The new materials can potentially foster polar mesospheric cloud and polar stratospheric cloud nucleation, allow new paths of heterogeneous chemistry in the stratosphere that could catalytically destroy ozone, influence radiative forcing and produce particles with unusual optical properties \citep{art_Schulz_2021, art_Boley_2021, art_Shutler_2022, art_Murphy_2023, techrep_Mitchell_2024}. There is therefore a wide variety of possible effects that could induce long-term changes of the atmosphere. If mass is ablated above 80\,km, influences and contribution to the ionosphere and even magnetosphere are possible \citep{art_Yamauchi_2024}. Modeling studies have started to shine light on the implication of the atmospheric injection, focusing on effects of future elevated aerosol emissions \citep{art_Jain_2025, art_Maloney_2025}, and giving insights into the molecular scale demise process \citep{art_Ferreira_2024}. Other studies have addressed the issue of space waste matter injection together with exhaust from rocket launches \citep[e.\,g.][]{art_Miraux_2022a, art_Ryan_2022, inPro_Fischer_2024}, and tried to work out a complete life cycle assessment of space activities \citep[e.\,g.][]{art_Miraux_2022}. 

Beside the space waste injection estimation that addressed 10 different elements by \citet{art_Schulz_2021}, other estimates have mainly focused on the aluminum input \citep{art_Ferreira_2024, art_Barker_2024,inPro_Fischer_2024}. Earlier studies \citep{techrep_ATISPADE_2019, techrep_ARA_2019} give more elaborate injection estimates but the origin of the data is unclear. The first observations of space waste remnants in the stratosphere \citep{art_Murphy_2023} have confirmed many of the estimates made in \citet{art_Schulz_2021}, such as the relative input of Al to Cu, the wide variety of materials being injected into the atmosphere, and the anthropogenic fraction of Al. However, there are also considerable differences and some metals detected by \citet{art_Murphy_2023}, were not analyzed in the injection estimates. Also, the past few years have already shown a significant increase in launch activity: there are nearly 10,000 large constellation satellites in orbit. With the first significant batches of constellation satellites being launched in 2020, and a typical lifetime of 5 years, we are now seeing the start of frequent reentries of these constellation satellites, in addition to the already high influx numbers. 

This study presents an update to the atmospheric injection of space waste presented by \citet{art_Schulz_2021}. Injection estimates for the years 2015 to 2025, as well as two future scenarios for 43 elements, are determined by using reentry databases, extending rocket stage knowledge and improving the estimation of the average elemental composition of satellites and rocket stages. To show the significance of the anthropogenic injection as well as to highlight essential future research paths, the resulting estimates are compared to both the natural influx and analyses of stratospheric aerosol particles \citep{art_Murphy_2023}.

\section{Anthropogenic injection}
As in \citet{art_Schulz_2021}, we estimate the space waste injection into Earth's atmosphere from the mass influx to the top of the atmosphere, the space waste elemental composition, and the ablation fraction. By combining this information, insights into the total and elemental annual injection rates are obtained. 

\subsection{Mass influx}
\label{sec:mass_influx}

\subsubsection{Data origin}
The space waste mass influx can be divided into orbital and suborbital objects reentering the atmosphere. Nearly all orbital objects are either satellites or rocket upper stages; a comparably small amount of mass is debris such as payload adapters, solar panel covers or disintegration remnants. We group them together with satellites. The annual mass influx of satellites and upper stages is obtained from the Catalog of Space Object Reentries of ``Jonathan's Space Report'' \citep{mics_McDowell_rcat_2025}, where we exclude all entries labeled suborbital (tagged D-SO), having left Earth's sphere of influence (U-HEL and D-LUN) and designed to land (D-LAND). The latter for example include manned return capsules (Soyuz-MS, Crew Dragon, Shenzhou, Orion), reconnaissance satellites (Yantar-4K2M) or unmanned return vehicles (X-37B, Chongfu Shiyong Shiyan Hangtian Qi, Cargo Dragon 2, Shi Jian 19). The upper stages of test flights 2, 7, and 8 of SpaceX's Starship are added due to their large mass reaching close to orbital velocity and observations of their destructive reentry. While there is a generally good agreement between \citet{mics_McDowell_rcat_2025} and ESA's DISCOS database \citep{mics_Discos}, we opt for the former as more realistic dry masses are provided and additional objects (such as separated service modules of reentry vehicles) are included. 

Suborbital objects are mainly rocket core stages which are jettisoned before reaching orbit. Thus, they reenter the atmosphere directly after their ascent with speeds ranging from 2 (when jettisoned early-on) to 8\,km/s (close to orbital speed). If the velocity is high enough, they too ablate upon re-entry. We adopt a threshold velocity of 3.8\,km/s from \citet{art_Schulz_2021}, which corresponds to about a quarter of the kinetic energy of a typical orbital reentry. Below that, we expect no significant ablation and thus disregard all booster stages (strap-on rocket boosters), nearly all payload fairings, and all purely suborbital launches (such as ICBM tests or sounding rockets). Orbital rocket launches are obtained from DISCOS \citep{mics_Discos} with the dry masses gathered from \citet{art_Barker_2024, art_Schulz_2021, misc_GuntersSpacePage, mics_McDowell_gcat_2025}, or estimated using launch data. Reentry velocities at 100\,km altitude are either taken from \citet{art_Schulz_2021}, calculated from trajectory data, or determined from launch coverage. A detailed list of the relevant core stages and their typical reentry velocities is provided in \hyperref[AppA]{Appendix A}.

\subsubsection{Historical reentry data (2015--2025)}
\begin{figure}
\centering
\includegraphics[scale=0.65]{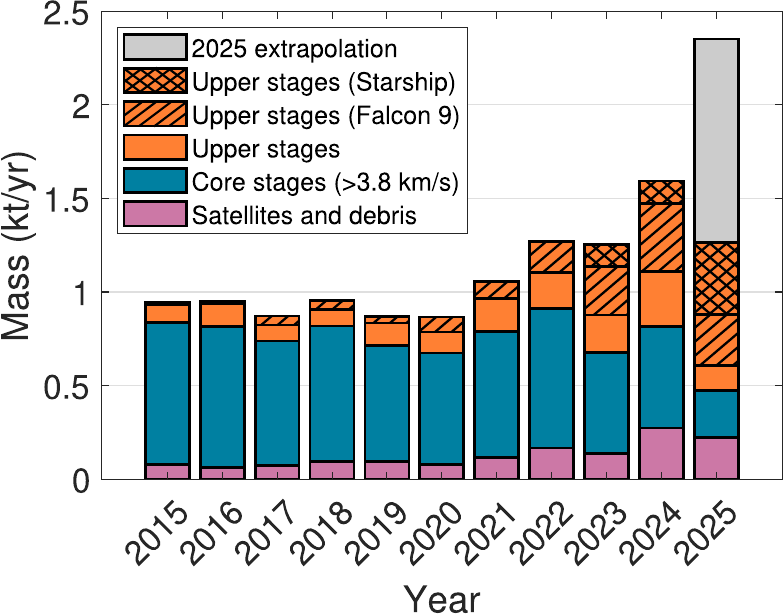}
\caption{Annual mass influx to the top of the atmosphere, with differentiation of object type. The gray bar shows the mass influx expected for the rest of 2025 when extrapolating the current numbers (until July 16th 2025) to the end of year.}
\label{fig:massinflux_type}
\end{figure}

Using the above described datasets, we determine the annual mass influx to the top of the atmosphere (prior to ablation) for the years 2015 to 2025 (until 16th July, 2025). We differentiate between satellites, rocket upper stages and (suborbital) core stages in order to account for their different composition and ablation fraction. Additionally, we divide upper stages into three categories --- Falcon 9/Falcon Heavy, Starship, and all others --- because of additional composition data available and significant mass contributions in recent years for the first two (especially Starship with 120\,t to 127\,t of mass per stage depending on the version). The mass influx is depicted in Figure \ref{fig:massinflux_type} and Table \ref{tab:mass_influx}, including an extrapolation to the end of 2025 assuming the same rate of influx as earlier in the year. While the annual mass influx has been more or less constant just below 1\,kt/yr for the years 2015 to 2020, with the onset of the first large satellite constellations, it has increased since 2020 to 1.6\,kt in 2024 and (considering the extrapolation) above 2.3\,kt in 2025. It is quite clear that the increase is mainly attributable to the increased large satellite constellation activity, mainly due to Starlink. This can be seen from the continuous increase in Falcon 9 upper stages reentering annually (they are often used to put Starlink satellites into orbit) and also the increasing number of reentering Starlink satellites \citep{art_Oliveira_2025}. With their typical lifetime of 5 years, we are just at the start of drastically increased reentry numbers of large constellation satellites.

\begin{table*}[]
\centering
\caption{Annual space waste mass influx to the top of the atmosphere (meaning before ablation), with differentiation of object type. All values in kt. Uncertainties are calculated as described in Section \ref{sec:errors}.}
\label{tab:mass_influx}
\begin{tabular}{lllllll}
\hline
Year                         & \begin{tabular}[c]{@{}l@{}}Satellites \\ and debris \end{tabular} & \begin{tabular}[c]{@{}l@{}}Core \\ stages\end{tabular} & \begin{tabular}[c]{@{}l@{}}Upper \\ stages\end{tabular} & \begin{tabular}[c]{@{}l@{}}Upper stages (only \\ Falcon 9/Heavy)\end{tabular} & \begin{tabular}[c]{@{}l@{}}Upper stages \\ (only Starship)\end{tabular} & Total \\ \hline
2015                           & 0.08 & 0.76 & 0.11 & 0.01 & 0.00 & 0.94 $\pm$ 0.07\\
2016                           & 0.06 & 0.75 & 0.14 & 0.01 & 0.00 & 0.95 $\pm$ 0.07\\
2017                           & 0.07 & 0.67 & 0.13 & 0.05 & 0.00 & 0.87 $\pm$ 0.06\\
2018                           & 0.09 & 0.72 & 0.14 & 0.05 & 0.00 & 0.96 $\pm$ 0.07\\
2019                           & 0.09 & 0.62 & 0.16 & 0.04 & 0.00 & 0.87 $\pm$ 0.06\\
2020                           & 0.08 & 0.60 & 0.19 & 0.08 & 0.00 & 0.87 $\pm$ 0.06\\
2021                           & 0.11 & 0.68 & 0.27 & 0.09 & 0.00 & 1.06 $\pm$ 0.07\\
2022                           & 0.17 & 0.75 & 0.36 & 0.17 & 0.00 & 1.27 $\pm$ 0.09\\
2023                           & 0.14 & 0.54 & 0.58 & 0.26 & 0.12 & 1.26 $\pm$ 0.09\\
2024                           & 0.27 & 0.54 & 0.78 & 0.36 & 0.12 & 1.59 $\pm$ 0.11\\
2025 (until Jul. 16th)         & 0.22 & 0.25 & 0.79 & 0.27 & 0.38 & 1.26 $\pm$ 0.09\\
2025 (whole   year, extrapol.) &      &      &      &      &      & 2.35 \\ \hline
\end{tabular}
\end{table*}

\subsubsection{Future scenarios}
Since 2020, the number of proposed satellites in large constellations has increased from 110,000 \citep{art_Schulz_2021} to more than a million \citep{art_Falle_2023}. Although a large portion of the currently proposed large satellite constellations is unlikely to be realized \citep{art_Falle_2023}, the actual number of launched satellites in large constellations rose unprecedentedly from zero in 2019 to more than 9,000 in 2025 \citep{mics_McDowell_satconst_2025}. At present these are nearly all either Starlink or OneWeb. The Chinese constellations Guowang and Qianfan as well as Amazon's Kuiper constellation are starting to be launched, with tens of satellites each already in orbit \citep{mics_EncyclopBrit_Megaconstellations}. Kuiper plans to deploy 3,000 satellites, Guowang about 13,000 and Qianfan around 15,000. Thus, the probable future scenario (Scenario 1) of 19,400 satellites in active constellations presented by \citet{art_Schulz_2021} might well be reached by 2030. If the current growth of the space industry and especially large satellite constellations is sustained for a few more decades to come, the worst-case future scenario (Scenario 2) of \citet{art_Schulz_2021} with 75,000 active constellations satellites is well within reach, too. Thus we adopt these two future scenarios here with 2.7 and 8.1\,kt of annual mass influx, respectively.

\subsection{Composition}
The composition of satellites and rocket bodies is (in many cases) proprietary knowledge of manufacturers and operators. Therefore, detailed composition information about specific launch vehicles or spacecraft is hard to obtain. Here, we combine information from space material databases with anonymized manufacturer data, institution and industry reports, and available scientific literature, to obtain a fairly complete picture of components and materials and their relative mass fractions. Through datasheets and other literature, and stoichiometry of used chemicals, the elemental mass composition of each component or material is determined. Combination of all the acquired information yields average elemental mass compositions of satellites, the different upper stages and core stages, given in \hyperref[AppB]{Appendix B}. These are significantly improved compared to the rough estimates presented by \citet{art_Schulz_2021}. This allows for credible mass fractions for a high number of elements besides the commonly considered aluminum \citep[e.\,g.][]{art_Barker_2024, inPro_Fischer_2024, art_Ferreira_2024}.

The main sources we have used are ``SPACEMATDB --- Space Materials DataBase'' \citep{mics_spacematdb}, which combines data from several sources, and the Aerospace Structural Metals Handbook \citep{book_ASMH_2007}. They include information about typical alloys and other materials used, datasheets, and general application areas of the materials as well as usage in certain launch vehicles and spacecraft.

\subsubsection{Satellite composition}
\begin{table*}[]
\centering
\caption{Estimated average composition of a typical LEO satellite. The mass fractions refer solely to the respective material groups, not elemental composition. Typical materials used for modeling of an average satellite composition are given in the fourth column.}
\label{tab:sat_comp}
\footnotesize
\begin{tabular}{p{3cm}p{1.5cm}p{3cm}p{4cm}p{4.5cm}}
\hline
Material groups                                                     & Mass fraction (\%) & Source of mass fraction                                                  & Typical materials                                                                                          & References                                                                                                          \\ \hline
Aluminum and aluminum alloys                                   & 36.7              & Mean                                                                      & AA7075, AA2024, AA6061, AA5052                                                                             & LSI list; \citet{art_Bonvoisin_2023, mics_spacematdb, techrep_Lerch_2018, man_AA7075, man_AA6061} \\
CFRP                                                           & 15.1              & Mean                                                                      & Carbon fibre + epoxy resin or cyanate ester resin                                                          &  \citet{art_Sun_2015, art_Bonvoisin_2023, art_Metzler_2016, art_Wen_2024, phd_Meier_2017, mics_Daken_Chemical}                                                                                                                         \\ 
Copper and copper alloys (not including copper in other materials)                                    & 1.4               & LSI                                                                  & Cu, Bronze (CDA 510), Ni-Al-bronze (CDA 632, C95500), Brass (CDA 230, CDA 260),  CuBe with coating of Au or Ag                      &  LSI list; \citet{mics_spacematdb, mics_C95500, mics_CuBe}   \\
Nickel and nickel alloys                                       & 0.5               & Mean                                                                      & Ni, Invar                                                                                                  &    LSI list; \citet{mics_spacematdb}                                                                                                                        \\
Titanium and titanium alloys                                   & 4.6               & Mean                                                                      & Ti6Al4, Ti3Al2.5V                                                                                          &   LSI list; \citet{mics_spacematdb, art_Bonvoisin_2023, man_Ti6Al4V}                                                                                                                         \\
Steels                                                         & 0.1               & LSI                                                                  & 4130, 4340                                                                                                 &  \citet{mics_spacematdb, mics_4340}                                                                                                                         \\
Stainless steels                                               & 3.2               & Mean                                                                      & 431, 301, 304L, 316L, A286, 17-4PH                                                                         & LSI list, ESA list, \citet{mics_spacematdb, man_304steel, man_A286, man_17_4PH}                                                                                                                          \\
Welding, brazing, soldering                   & 1.1               & Mean                                                                      & SnPb (Solder)                                                                                                      & LSI list, \citet{mics_Kester_solder}                                                                                                                          \\
Miscellaneous metallic materials                             & 0.4               & Mean                                                                      & AZ31B                                                                                                      &     \citet{mics_spacematdb}                                                                                                                      \\
Optical materials                                              & 2.9               & Mean                                                                      & BK7, ZERODUR, SiO2                                                                                         &   LSI list,  \citet{mics_Zerodur, mics_BK7}                                                                                                                        \\
Adhesives, coatings, varnishes                               & 5.4               & LSI                                                                  & Acrylate, Silicone, Epoxy                                                                                  &    LSI list, \citet{mics_spacematdb,man_acrylate_Loctite, man_Silicone_resin, man_Epoxy_adhesive_Scotchweld, man_Epoxy_adhesive}                                                                                                                      \\
Adhesive tapes                                                 & 1.6               & LSI                                                                  & Acrylate                                                                                                   &   LSI list                                                                                                                          \\
Paints, primers and inks                                       & 0.5               & LSI                                                                  & Polyurethane, Silicone + ZnO                                                                               &  LSI list, \citet{mics_spacematdb, man_black_paint_Aeroglaze, art_LecadreScotto_2025, art_Kayhan_2012}                                                                                                                         \\
Lubricants                                                     & $<0.1$               & Mean                                                                      & MoS2, PFPE                                                                                                 &    LSI list, \citet{mics_spacematdb}                                                                                                                      \\
Potting compounds, sealants, foams                           & 1.9               & LSI                                                                  & Polyurethane, Silicone, Epoxy                                                                              &   \citet{mics_spacematdb, man_EccostockFPH_PU, man_CV1143, art_Buschow_2001, man_epoxy_foam_Stycast}                                                                                                                        \\
Rubbers and elastomers                                       & 0.2               & Mean                                                                      & Viton B910, Nitrile Butadiene Rubber                                                                       &   LSI list, \citet{mics_spacematdb, pat_Barber_2001, mics_NBR}                                                                                                                            \\
Thermoplastics (non-adhesive tapes and foils, e.\,g. MLI)				& 3.7               & LSI                                                                  & Polyimide, Polyolefin, DACRON, PTFE, PEEK, PVDE                                                            &   \citet{mics_spacematdb, art_Shu_2016, man_Kapton}                                                                                                                        \\
Wires and cables                                             & 7.2               & Mean of LSI and \citet{techrep_ATISPADE_2019}                                                  & Cu, PTFE, PFA, Polyimide, Ag                                                                               &    \citet{man_Gore_2016}                                                                                                                       \\
Miscellaneous non-metallic materials (e.\,g. ceramics)          & 1.0               & LSI                                                                  & Al2O3, Si3N4, AlN, SiC                                                                                     &   LSI list, \citet{art_Bonvoisin_2023}                                                                                                                        \\
Electronics (PCBs)                                      & 5.0               & \citet{book_NewSMAD_2011}                                                                      & Polyimide, FR4, electronic elements                                                                        &   \citet{art_Korf_2019, art_Mumby_1989, mics_spacematdb}                                                                                                                        \\
Li-Ion Batteries (whole)                                       & 6.4               & Calculated using \citep{book_NewSMAD_2011}, other sources (see text)                                     & Al, Cu, LiPF6, LiCoO2, LiNi0.84Co0.12Al0.04O2, Graphite, C2H4                                              &  \citet{InCol_Ehrlich_2018, inCol_BORTHOMIEU_2014, art_Pathak_2023, man_LFC40, man_Yuasa, inPro_Bonneau_2017, man_VES16_2020}                                                                                                                         \\
Solar panels (solar cells + cover glass)                          & 1.0               & Calculated assuming 10\,m$^2$ of solar array & CMX100 cover glass, Triple junction solar cells substrate (Ge + GaAs, GaInP), silicon solar cell substrate &   \citet{man_Azur1_2016, man_Azur2_2016, man_AzurSi_2016, man_Sharp_2023, art_Lin_2016, art_Li_2021, man_QIOPTIQ, art_Kearsley_2007}                                                                                                                        \\ \hline
Total                                                            & 100.0             &                                                                           &                                                                                                            &                                                                                                                           \\ \hline
\end{tabular}
\end{table*}

Satellites re-entering Earth's atmosphere are predominantly from LEO. A large European satellite manufacturer (LSI) provided an anonymized dataset, which lists the mass fraction of material groups \citep[according to the ECSS-Q-ST-70 material classification,][]{techrep_ECSS} of a typical LEO satellite ($<$1000\,kg in mass) based on a metal structure. We combine this data with mean satellite compositions given in the ATISPADE \citep{techrep_ATISPADE_2019} and ARA studies \citep{inPro_ARA_2017, techrep_ARA_2019}, as well as in \citet{inPro_Beck_2019}. For these studies, single materials are provided rather than material groups (e.\,g. copper instead of its material group copper alloys). We allocate the materials to the appropriate material groups if possible, in order to reach mass fractions for each material group. Depending on the maturity of the data, we chose a mean value of all studies or took values from only a subset of studies to reach an average mass fraction for each material group. Representative materials of each group are then found by combining data provided by the LSI, an ESA (European Space Agency) materials list, \citep{mics_spacematdb} and datasheets. 

The given material groups do not include (or only partially include) components like PCB (printed circuit board) electronics, batteries, solar cells, and CFRP (carbon fibre-reinforced polymer). As these components are significant contributors especially regarding specific metallic elements, we estimate them separately. While CFRP can be estimated using the data from all 4 datasets, the other components are not listed consistently and the values that are presented seem to be zero-order estimates with unlikely values. Thus, different sources are used to assess the mass fraction of these components. The final mass fractions are obtained by renormalizing all other obtained mass fraction values so that a 100\% total is reached. Table \ref{tab:sat_comp} shows the material groups/components together with the associated mass fraction, the source of the mass fraction and typical, representative materials. In the following, the components respectively material groups will be discussed individually. 

\paragraph{Aluminum, aluminum alloys and CFRP}
Aluminum (Al) alloys are predominantly used in structures, with CFRP gaining more and more ground. The link between structural application and mass fraction can also clearly be seen from the LSI data, which shows no or very little use of CFRP while the typical satellite assumed by \citet{techrep_ATISPADE_2019, inPro_Beck_2019} attributes 25\% of the satellite mass to CFRP. This reduces the Al alloy portion from 54\% in the LSI data to 27 and 35\% in \citet{techrep_ATISPADE_2019} and \citet{inPro_Beck_2019}, respectively. The values in \citet{inPro_ARA_2017} are in between, for both CFRP and Al alloys. For both Al alloys and CFRP, we have taken the average value of all 4 studies.

CFRP is made of carbon fibres which are embedded in an exopy or cyanate ester resin. Typically, the carbon fibres make up about 60\% of the material \citep{art_Sun_2015, art_Bonvoisin_2023}. For epoxy resin based CFRP, we use the mean of XPS analysis performed by \citet{art_Sun_2015, art_Metzler_2016, art_Wen_2024} assuming volume fractions to be reasonably close to mass fractions and accounting for the missing hydrogen with 3.2\% of the total mass. Besides the usual organics, a small portion of chlorine ($<$1\%) remains in the epoxy resin due to incomplete synthesis of epichlorohydrine. For cyanate ester resins, we assume bisphenol A dicyanate in a matrix together with the typical triazin repeating unit \citep{phd_Meier_2017, mics_Daken_Chemical}. Stoichiometry yields the elemental mass composition.

\paragraph{Other alloys and metals}
Copper, nickel, titanium and their alloys as well as steels, stainless steels and other metallic elements are found in various different applications such as  structural elements, fuel tanks, electronic assemblies, nozzles and thrusters, actuators, bearings, thermal systems, plating and other applications. For details on the applications, the reader is referred to \citet{mics_spacematdb, InCol_Finckenor_2018}. Most common alloys and metal types in use are obtained from the material list from the LSI, the ESA materials list and \citet{mics_spacematdb}. The respective alloy composition is either given in the mentioned references or in industry datasheets.

\paragraph{Filler metals, welding, brazing, soldering}
Here, solder is the major component. While efforts are being made to move to lead-free solder, typical solder with 40\% Pb and 60\% Sn is still widely in use \citep{mics_spacematdb, man_Elsold}.

\paragraph{Optical materials}
Typical glasses used are BK7 and ZERODUR. These consist mainly of silica (SiO$_2$) combined with other metal oxides such as B$_2$O$_3$, ZnO, Al$_2$O$_3$, Na$_2$O, TiO$_2$, K$_2$O, ZrO$_2$, MgO, P$_4$O$_{10}$, Li$_2$O, BaO and trace amounts of  CaO and As$_2$O$_3$. Approximate compositions are given in \citet{mics_BK7, mics_Zerodur}. Fused silica glass consisting purely of silica is used as well. We assume equal mass fraction of these three kinds of glasses. Solar cell cover glass is addressed under solar panels below.

\paragraph{Adhesives, coatings, varnishes and adhesive tapes}
We group these into acrylate, silicone and epoxy based, assuming equal mass fraction for each type. Approximate chemical compositions of representative substances are available in \citet{man_acrylate_Loctite} for an acrylic adhesive, in \citet{man_Silicone_resin} for a silicone varnish, and in \citet{man_Epoxy_adhesive_Scotchweld, man_Epoxy_adhesive} for two epoxy adhesives; the latter being electrically conducting due to a high silver content. For the epoxy resin itself, we assume Bisphenol A diglycidyl ether, as it is commonly used for epoxy resins. As with the CFRP epoxy resins, we assume proportionate trace amounts of chlorine in the epoxy resin due to incomplete synthesis of epichlorohydrine.

For adhesive tapes, no chemical composition of a representative product could be found. As acrylate is used, we assume the same composition as for the acrylate adhesive.

\paragraph{Paints, primers and inks}
Both polyurethane (PU) and silicone based paints are used. We chose a black, PU based paint and a white, silicone based paint as these are commonly used \citep{mics_spacematdb} as both heating and cooling may be needed in the LEO environment. A representative chemical composition of a black paint with black carbon as pigment is obtained from \citet{man_black_paint_Aeroglaze}, assuming the ester solvents as butyl acetate. White paints normally contains ZnO as the pigment; we assume polydimethylsiloxane (PDMS) as the silicone binder \citep{art_LecadreScotto_2025} and the ZnO mass fraction to be 27\% \citep{art_Kayhan_2012}. For both, the elemental mass composition is determined through stoichiometry.

\paragraph{Lubricants}
While lubricants only make up a very small overall mass fraction of a satellite, we model them as the dry lubricant MoS$_2$ and perfluoropolyether (PFPE). The elemental mass compositions are determined via the stoichiometric ratios, assuming a reapeating unit of C$_2$F$_4$O for PFPE.

\paragraph{Potting compounds, sealants, foams}
Taking acount of the materials described in \citet{mics_spacematdb}, we assume an equal share of PU foam, epoxy and silicone based materials. Again, representative materials are chosen from datasheets. For PU foam, we model the composition after \citet{man_EccostockFPH_PU} with 60\% Toluene-2,4-Diisocyanate (TDI) and the rest an isocyanate terminated polymer (dipropylen glycol). For silicone, we chose \citet{man_CV1143}, which contains 20\% silica and 15\% 2-Butanone, 0,0',0" (methylsilylidyne) trioxime, the rest is assumed to be silicon binder (PDMS, see above). Epoxy foam is modeled after \citet{man_epoxy_foam_Stycast}, assuming E-glass being used \citep[e.\,g.][]{art_Buschow_2001} and trace amounts of chlorine remaining from incomplete synthesis of epichlorohydrine (as above). 

\paragraph{Rubbers and elastomers}
Viton B910 as well as nitrile butadiene rubber (NBR) are used. The former consists of vinylidene fluoride (1,1-Difluoroethylene) and a hexafluoropropylene copolymer \citep{mics_spacematdb} with the latter making up about 10\% of the mass \citep{pat_Barber_2001}. NBR is derived from acrylonitrile (ACN) and butadiene, with ACN making up about 34\% of the mass \citep{mics_NBR}. Thus, both materials' elemental composition can be calculated using stoichiometry.

\paragraph{Thermoplastics (e.\,g. MLI)}
Here, we assume 70\% of the mass being multi-layer insulation (MLI) with representative materials polyimide \citep[namely Kapton, which is poly(4,4'-oxydiphenylene pyromellitimide, ][]{art_Shu_2016}, polyolefin (such as polyehtlyene (PE) or polypropylene (PP)) and polyethylene terephthalate (PET). The optical properties of MLI are normally achieved by a thin aluminized layer. Based on its thickness relative to the thickness of the whole sheet, we 
calculate the aluminum content to 0.03\% \citep{man_Kapton}. The remaining 30\% of the mass are attributed to polytetrafluoroethylene, polyether ether ketone and polyvinylidene fluoride.

\paragraph{Miscellaneous non-metallic materials}
Here, ceramics such as alumina (Al$_2$O$_3$), silicon nitride (Si$_3$N$_4$), aluminum nitride (AlN) and silicon carbide (SiC) \citep{art_Bonvoisin_2023} are the dominating mass contributors. Stoichiometry yields the elemental mass composition.

\paragraph{Batteries}
Today's satellites almost exclusively use Li-ion batteries as they are superior to older battery types in various ways, e.g. specific energy and power density \citep{inCol_BORTHOMIEU_2014, art_Pathak_2023}. Two types, namely LCO (LiCoO$_2$) and NCA (LiNi$_{0.84}$Co$_{0.12}$Al$_{0.04}$O$_2$), are used for the positive electrode \citep{inCol_BORTHOMIEU_2014, man_LFC40, man_Yuasa, inPro_Bonneau_2017, man_VES16_2020}, while the negative electrode is normally graphite, along with a LiPF$_6$ electrolyte and polyolefin separator \citep{InCol_Ehrlich_2018}. The relative mass contribution of the different constituents is known \citep{InCol_Ehrlich_2018}. We assume the casing to be aluminium, the cap assembly to be copper, and the separator to be PE. Considering typical capacities needed for LEO satellites \citep[100-150\,Ah, cf.][]{book_NewSMAD_2011} and available datasheets of batteries for use in the LEO environment \citep{man_VES16_2020,man_Eaglepicher_2022}, battery mass is about 25 to 40\,kg. Assuming a typical LEO satellite mass of about 600\,kg, this translates to 4.0 to 6.4\% of the satellite mass in batteries. Considering values of roughly 7.5\%  and 9.1\% derived from mass budget information in \citet{techrep_Dudley_1997}, and  \citet{book_NewSMAD_2011}, respectively, we adopt the upper limit of 6.4\%.

\paragraph{Electronics (PCBs)}
According to \citet{book_NewSMAD_2011}, the on-board computer generally makes up 5\% of the mass of a satellite. We assume this to be only PCB mass. For its composition, we take the mean PCB composition from the literature review in \citet{art_Korf_2019} and amend the data by the organic elements not analyzed (H, C, O). Those are determined from the elemental mass fractions of a FR4 PCB laminate material \citep{art_Mumby_1989}, which is still commonly used today \citep[e.\,g.][]{mics_spacematdb}.

\paragraph{Wires and cables}
For wiring and cables, a mean mass fraction of around 7 to 8\% of the satellite mass is obtained LSI and \citet{techrep_ATISPADE_2019}, consistent with the copper content given in \citet{inPro_Beck_2019}. While the majority of the material of a cable is copper, a substantial mass portion is insulation and jacket. Trace amounts of silver are used for better conduction. We use a data sheet from a major space cable distributor \citep{man_Gore_2016} giving detailed information about the thicknesses of the copper core, the very thin silver coating as well the insulation and jacket, which normally consist of PTFE, polyimide and PFA. We chose several shielded and non-shielded SPM cables and find an average of the copper (just above 50\%), silver (below 1\%) and plastic content. 

\paragraph{Solar panels}
We estimate the solar array size of a typical LEO satellite to be around 10\,m$^2$, based on the Iridium NEXT satellites \citep[e.\,g.][]{mics_Iridium_SA}. Solar arrays consist of the structure, wiring, the solar cell itself as well as cover glass. The former two are modeled separately (see above). The actual solar cells are extremely thin, as is the cover glass. Datasheets give values of the density of the whole assembly between 32 to 150\,mg/cm$^2$ \citep{man_Azur1_2016, man_Azur2_2016, man_AzurSi_2016, man_Sharp_2023}, which translates to an average of 9.1\,kg of solar cell and cover glass --- about 1\% of the satellite's mass. Appropriate values are 65\% of the mass being solar cell material, and 35\% cover glass. 

Due to their superior efficiency, triple junction InGaP/GaAs/Ge solar cells are used much more frequently than the cheaper silicon-only cells. Based on the availability of datasheets and the present product palettes, we assume 80\% being multi-junction solar cells. These exhibit a complex layering of differently doped substances with the majority of the mass being the Ge substrate. We model this layering after \citet{art_Lin_2016, art_Li_2021} and thus determine the elemental composition. For both cells, a silver layer is used for electric contact. The cover glass is normally borosilicate glass doped with cerium dioxide to shield from UV radiation  \citep[e.\,g. CMX, ][]{man_QIOPTIQ, art_Kearsley_2007}). We model it after the cover glass from the Hubble space telescope (pers. comm. A. T. Kearsley, Natural History Museum, London, UK).

\subsubsection{Average upper stage composition}
We determine a typical upper stage composition by averaging component mass fractions of the AVUM stage of the Vega rocket \citep{inproceedings_Dumon_2022} and the Delta II upper stage \citep{inPro_Lips_2013, techrep_Steckel_2018}. The data is complemented by information obtained from \citet{inPro_Tizon_2017, techrep_ATISPADE_2019, inPro_Beck_2019}. Again, after averaging, we renormalize the total component mass fraction to 100\%. Solid rocket motors are neglected as they are less commonly used in upper stages. In any case, they would not introduce different materials compared to liquid fueled rocket stages: paints and reaction control system components are similar, common steels and CFRP are used for structural components and nozzles, and insulation fibres and nozzle ablators are compositionally similar to CFRP \citep{InCol_Henson_2018, inBook_Trinh_2018}.

Common materials of rocket upper stages are gathered from \citet{InCol_Halchak_2018, InCol_Henson_2018, mics_spacematdb} as well as the Aerospace Structural Metals Handbook \citep{book_ASMH_2007}, datasheets and other scientific literature. While there are obvious strong differences in the construction of rocket stages compared to satellites, many of the components and materials used in satellites are also used in rocket stages, like paint, MLI, batteries, wires, solder, and PCBs. These components are modeled as for the satellites. The resulting mass composition as well as the relevant materials in use are given in Table \ref{tab:RB_comp}. In the following, the different components and their typical materials are described.

\paragraph{Propellant tanks, structure and payload adapters}
They make up the largest mass portion by far. Aluminum alloys are favored due to their low weight \citep{InCol_Henson_2018}. The workhorse alloy is 2219 (we assume 50\%), but 2014 and newer aluminum lithium alloys like 2195 are used as well (each assumed 10\% mass). Stainless steels are used to a lesser extent, for example in rare (because technically challenging) ballon tank designs such as 301 in the Centaur upper stage. But they appear also in standard tank designs such as 410  used in the Delta II upper stage \citep{techrep_Steckel_2018}; we assume 5\% each. For structure, aluminum alloys like 7075 as well as CFRP are used \citep{inproceedings_Dumon_2022, inPro_Beck_2019, techrep_ATISPADE_2019}. We assume 10\% mass portion for each. 

Payload adapters normally are made from aluminum alloys such as 7075 as well as CFRP \citep[e.\,g.][]{inproceedings_Dumon_2022}; we assume an equal share of mass.

\paragraph{Rocket engines}
Rocket engines comprise several components, namely a thrust chamber inner liner, thrust chamber jacket, nozzle and nozzle extension, feedlines and tubes, turbopumps, and valves. The mass fraction of each of these components differs depending on the engine cycle. We compute an average of the mass distribution of these components given in \citet{inPro_Tizon_2017} for the three different cycles. By multiplying these mass fractions with the average mass fraction of an engine in a rocket upper stage, one obtains the overall mass fraction of the specific engine components.

The thrust chamber wall (sometimes referred to as the inner liner) consists of alloys that have been specifically designed to handle the high loads and temperatures. These are specific copper and copper-chromium alloys like Narloy-Z and CRCop-84 \citep{InCol_Halchak_2018, book_Sutton_2017, techre_Ellis_2005}, or superalloys like CTX 1, 3 and 909 \citep{man_CTX1, man_CTX3, man_CTX909}, and nickel-chromium alloys such as X-750 \citep{man_X750}. We assume 20\% mass fraction each for Narloy-Z, CRCop-84, CuCr3, the CTX alloy class and X-750. 

For the thrust chamber jacket, which usually cools the inner thrust chamber walls by propellant running through its tubes, sometimes the same materials as for the inner liner are used \citep{book_Sutton_2017}. But stainless steels and nickel-chromium alloys are also common materials \citet{InCol_Halchak_2018}. Based on the frequency of use reported in the relevant literature, we assume 40\% mass fraction shared equally between 316 \citep{man_316steel}, 347 \citep{mics_347steel} and A286 \citep{man_A286} stainless steels, and 15\% shared equally between  Inconel 600 \citep{man_inconel600}, X-750, Narloy-Z and CRCop-84, each.

Nozzles can consist of the same materials as thrust chamber inner liners or thrust chamber jackets, but different materials are also sometimes employed \citep{InCol_Halchak_2018}, for example nickel-chromium alloys such as 718 \citep{ man_718alloy_carpenter,man_718alloy_boehler}, Ti6Al4V \citep{man_Ti6Al4V}, and C-103 Niobium alloy \citep{man_C103}. These are also used in rocket nozzle extensions which are needed for vacuum engine versions. Due to the generally lower temperatures compared to the nozzle close to the thrust chamber, CFRC is also a common option \citep{InCol_Halchak_2018, mics_RL10}. Based on the literature, we tentatively assume 25\% CRFC, 20\% C-103, and 20\% shared equally between the three stainless steels (above) and 10\% shared equally between 718, Inconel 600 and Ti6Al4V, each. 

Feedlines and tubes are usually constructed from austenitic stainless steels, mainly 321 \citep{man_321steel}, although 347 and A286 stainless steel, nickel-chromium 718 alloy, Hastelloy \citep{mics_spacematdb}, aluminum alloys such as 6061 \citep{man_AA6061} and Ti6Al4V are also used \citep{InCol_Halchak_2018, InCol_Henson_2018}. Based on their frequency of use as reported in the relevant literature, we tentatively assume 30\% 321 steel, 20\% each for 347 and A286 steel, 10\% each for Al6061 and Ti6Al4V,  and 5\% each for 718 and Hastelloy. 

Turbopumps employ a wide range of durable and temperature resistent materials \citep{InCol_Halchak_2018}. Housing components are composed of nickel-chromium alloys 625 \citep{man_inconel625} and 718, stainless steels like A286 and Incoloy 903 \citep{man_Incoloy903}, and cobalt-based alloys like Haynes 188 \citep{man_haynes188}. Pumping elements need extremely hard materials such as silicon nitride, Cronidur 30 \citep{man_Cronidur30} and 440C stainless steel \citep{man_440C}, or 9310 low-carbon steel \citep{man_9310}. While other alloys and steels are certainly in use, the materials classes will not be largely different due to the physical conditions the materials must endure. Therefore, we assume equal mass fractions of the 9 materials listed here to obtain a representative average turbopump composition. 

According to \citet{InCol_Halchak_2018}, valves usually consist of an aluminum alloy housing (like A357 \citep{man_A357}and 7075 alloys \citep{man_AA7075}), with nickel and titanium based alloys (such as Inconel 625 and Ti5Al2.5Sn \citep{man_Ti5Al25Sn}) used for high-strength needs. The valve internals favor specific steels such as 15-5PH steel \citep{man_155PH}. We assume a 20\% equal share of these 5 representative materials.

\paragraph{Pressurization tanks}
These tanks are made for pressurizing the main fuel tanks when their pressure drops due to more and more fuel being burnt by the engine. Composite overwrapped pressure vessel (COPV) tanks are common \citep[e.\,g.][]{inproceedings_Dumon_2022} as well as Ti6Al4V tanks \citep[e.\,g.][]{techrep_Steckel_2018, inPro_Lips_2013}to hold the pressurization gas (normally helium). COPV tanks commonly consist of a Ti6Al4V liner which is wrapped by a CFRP-epoxy layer \citep{man_MT_Aerospace_Tanks}. All in all, we asssume a mass composition of 50\% each of CFRP-epoxy and Ti6Al4V.

\paragraph{Reaction control system (RCS)}
An RCS system is needed to control spacecraft attitude. It consists of propellant tanks, made of Ti6Al4V \citep{man_MT_Aerospace_Tanks, man_Arianespace_PropTanks, techrep_Steckel_2018} and multiple small thrusters to allow direction changes. Thruster use a lot of different materials \citep{man_Arianespace_HydrTrhust, man_NammoThruster, InCol_Halchak_2018} like Haynes 25 \citep{man_haynes25}, 304L \citep{man_304steel} and 4550 \citep{man_Radiometal4550} steel, ethylene propylene diene monomer (EPDM) rubber \citep[determined from][]{art_Oliviera_2010} as well as carbon and glass cloth ablators \citep[calculated from][]{man_CarbonCloth, man_F502_PhenolicPrepreg, mics_EGlass_AZOM}. We assume a small portion (5\%) of ablator remains until reentry, 10\% EPDM rubber and the other 85\% mass an equal share of the alloys.

\paragraph{Guidance/Avionics}
Typically, electronics boxes made of aluminum alloys such as 6061 are used to house the PCBs. Other components include (as in satellites) wires, lithium-ion batteries and solder. We tentatively assume 50\% electronics box mass, 20\% PCB and wire mass (each) as well as 5\% of solder (as in satellites, about a quarter of the PCB mass) and 5\% of battery mass. The latter are assumed with a lower relative mass compared to satellites as especially for LEO launches, upper stages reenter within only a few orbits, thus no large battery charge is needed.

\paragraph{Others}
A significant mass fraction also goes into paint, insulation foam for cryogenic stages, and MLI. The compositions of these items are adopted from the ones calculated for satellites, where assuming only white paint (with zink oxide pigments) and PU foam as insulation foam.

\begin{table*}[]
\centering
\caption{Estimated average composition of different kinds of rocket stages.}
\label{tab:RB_comp}
\scriptsize
\begin{tabular}{lllllp{1.5cm}p{2.5cm}p{4.5cm}}
\hline
Component                                & \multicolumn{4}{c}{Mass fraction estimates (\%)}                                                                                                                        & Source of mass fraction\newline (upper stage) & Typical materials                                                                                                   & Material sources                        \\ \hline
                                                        & \begin{tabular}[c]{@{}l@{}}Upper \\stage \end{tabular}  & \begin{tabular}[c]{@{}l@{}}Core\\ stage\end{tabular}    & \begin{tabular}[c]{@{}l@{}}Falcon 9/Hvy. \\ upper stage\end{tabular} & \begin{tabular}[c]{@{}l@{}}Starship \\ up. st.\end{tabular} &                                       &                                                                                                                     &                                         \\ \hline
Rocket engine                                           & \textbf{9.6}  & \textbf{18.0} & \textbf{13.8}                                                         & \textbf{18.0}                                                   & Mean AVUM, Delta II                   &                                                                                                                     &  \\
                    \quad Feedlines/Tubes                     & 0.8           & 1.5           & 1.1                                                                   & 1.5                                                             & \citet{inPro_Tizon_2017}                            & 321, 347, A286 stainless steel, Al 6061, Ti6Al4V, Inconel 718, Hastelloy                                            &   \citet{InCol_Henson_2018,InCol_Halchak_2018,man_321steel, mics_347steel, man_A286, man_AA6061, man_Ti6Al4V, man_718alloy_carpenter, man_718alloy_boehler, mics_spacematdb}                                     \\
                     \quad Valves                            & 0.9           & 1.6           & 1.2                                                                   & 1.6                                                             & \citet{inPro_Tizon_2017}                              & Al A357, 7075, Inconel 625, Ti5Al2.5Sn, 15-5 PH steel                                                               &    \citet{InCol_Halchak_2018,man_A357, man_AA7075, man_inconel625, man_Ti5Al25Sn, man_155PH}                                     \\
                           \quad Thrust chamber inner liner  & 2.1           & 3.9           & 3.0                                                                   & 3.9                                                             & \citet{inPro_Tizon_2017}                              & Narloy-Z, Cu-Cr alloy, CRCop-84, CTX 1, CTX 3, CTX 909, HPM X750                                                    &                            \citet{InCol_Halchak_2018, book_ASMH_2007, man_CTX1, man_CTX3, man_CTX909, man_X750}             \\
                           \quad Turbopumps                  & 2.7           & 5.2           & 3.9                                                                   & 5.1                                                             & \citet{inPro_Tizon_2017}                              & Silicon nitride, Inconel 625, A286, Incoloy 903, Haynes 188, Cronidur 30 stainless steel, 440C, E 9310, Inconel 718 &                      \citet{InCol_Halchak_2018,man_inconel625, man_A286, man_Incoloy903, man_haynes188, man_Cronidur30, man_440C, man_9310,man_718alloy_carpenter, man_718alloy_boehler} \\
                            \quad Thrust chamber jacket       & 0.4           & 0.8           & 0.6                                                                   & 0.8                                                             & \citet{inPro_Tizon_2017}                              & 316, 347, A286, Inconel 600, HPM X750, Narloy-Z, CRCop-84                                                           &                \citet{InCol_Halchak_2018,man_316steel, mics_347steel, man_A286, man_inconel600, man_X750, book_ASMH_2007}                         \\
                           \quad Nozzle and nozzle extension & 2.7           & 5.0           & 3.8                                                                   & 5.0                                                             & \citet{inPro_Tizon_2017}                              & C-103, 316, 347, A286, Inconel 718, Inconel 600, Ti6Al4V, CFRC, Cu-Cr alloy                                         &                          \citet{InCol_Halchak_2018,man_C103,man_316steel, mics_347steel, man_A286, man_718alloy_carpenter, man_718alloy_boehler, man_inconel600, man_Ti6Al4V, mics_RL10}               \\
Pressurization tanks                                    & 5.2           & 5.1           & 5.0                                                                   & 2.9                                                             & Mean AVUM, Delta II                   & Ti6Al4V, CFRP (epoxy-based)                                                                                         &     \citet{man_MT_Aerospace_Tanks,man_Ti6Al4V, art_Sun_2015, art_Metzler_2016, art_Wen_2024, art_Bonvoisin_2023}                                    \\
RCS system                                              & \textbf{3.8}  & \textbf{3.8}  & \textbf{3.6}                                                          & \textbf{2.2}                                                    &                                       &                                                                                                                     &                                         \\
                           \quad Monopropellant tanks        & 1.6           & 1.6           & 1.6                                                                   & 0.9                                                             & Mean AVUM, Delta II                   & Ti6Al4V                                                                                                            &    \citet{man_MT_Aerospace_Tanks, man_Arianespace_PropTanks, man_Ti6Al4V}                                      \\
                           \quad Monopropellant thrusters    & 2.2           & 2.1           & 2.1                                                                   & 1.2                                                             & Half of AVUM                          & Carbon cloth phenolic, Glass cloth phenolic, Haynes 25, 304L, Radiometal 4550, EPDM rubber                          &                       \citet{man_NammoThruster, InCol_Halchak_2018, man_haynes25, man_304steel, man_Radiometal4550, art_Oliviera_2010, man_CarbonCloth, man_F502_PhenolicPrepreg, mics_EGlass_AZOM}
                                             \\
Guidance/Avionics                                       & \textbf{15.5} & \textbf{15.2} & \textbf{14.8}                                                         & \textbf{8.8}                                                    & Mean AVUM, Delta II                   &                                                                                                                     &                                         \\
                           \quad Electronics Boxes           & 7.7           & 7.6           & 7.4                                                                   & 4.4                                                             & Assumed 50\%                          & Al 6061                                                                                                             &   \citet{man_AA6061}                                      \\
                           \quad Cables/Wires                & 3.1           & 3.0           & 3.0                                                                   & 1.8                                                             & See satellite                         & See satellite                                                                                                       &  See satellite                                       \\
                           \quad PCBs                        & 3.1           & 3.0           & 3.0                                                                   & 1.8                                                             & See satellite                         & See satellite                                                                                                       &  See satellite                                       \\
                           \quad Solder                      & 0.8           & 0.8           & 0.7                                                                   & 0.4                                                             & See satellite                         & See satellite                                                                                                       & See satellite                                        \\
                           \quad Li-ion batteries            & 0.8           & 0.8           & 0.7                                                                   & 0.4                                                             & See satellite                         & See satellite                                                                                                       &  See satellite                                       \\
Payload adapter                                         & 7.0           & -           & 6.7                                                                   & -                                                             & Mean AVUM, Delta II                   & CFRP, AA 7075                                                                                                       &  \citet{inproceedings_Dumon_2022}                                       \\
Propellant tanks/Structure                              & 51.5          & 50.6          & 49.1                                                                  & 60.0                                                            & Mean AVUM, Delta II, \citet{techrep_ATISPADE_2019}         & AA 2219, AA 2014, AA 2195, 301, 410, CFRP, 7075, 2198 (Falcon upper stage), 304 (Starship)                          &    \citet{InCol_Henson_2018, techrep_Steckel_2018, inproceedings_Dumon_2022, inPro_Beck_2019, techrep_ATISPADE_2019, InCol_Wanhill_2014}                                     \\
Paint                                                   & 2.2           & 2.2           & 2.1                                                                   & -                                                             & \citet{techrep_ATISPADE_2019}, \citet{inPro_Beck_2019}                        & Silicone +  zinc oxide                                                                                              &   \citet{art_LecadreScotto_2025, art_Kayhan_2012}                                      \\
Insulation foam                                         & 2.5           & 2.5           & 2.4                                                                   & -                                                             & \citet{techrep_ATISPADE_2019}, \citet{inPro_Beck_2019}                         & Polyurethane                                                                                                        &  See satellite                                       \\
MLI/Thermal protection                                  & 2.7           & 2.6           & 2.5                                                                   & 1.5                                                             & \citet{techrep_ATISPADE_2019}, \citet{inPro_Beck_2019}                         & See satellite                                                                                                       &    See satellite                                      \\
Thermal protection system                               & -           & -           & -                                                                   & 6.7                                                             & Calculated (see text) & Silica fibres                                                                                                       &  \citet{mics_HRSItile}                                       \\ \hline
Total                                                     & 100           & 100           & 100                                                                   & 100                                                             &                                       &                                                                                                                     &                                         \\ \hline
\end{tabular}
\end{table*}

\subsubsection{Falcon 9/Heavy upper stage composition}
The Falcon 9/Heavy upper stage is modeled separately as it contributes a large fraction of the overall mass influx, as shown in Section \ref{sec:mass_influx}. Its Merlin 1D vacuum engine makes up 550\,kg of the approx. 4.25\,t dry mass of the stage \citep{mics_Wade_2019, art_Barker_2024}. As there is no detailed information about the remaining mass distribution available, we renormalize all other components of the average upper stage, so the total mass fraction is 100\%. The Falcon 9/Heavy upper stage is known to use 2198 aluminum lithium alloy for its propellant tank walls, while the domes are made from standard aluminum alloy (we assume a share of 2219 and 2014) \citep{InCol_Wanhill_2014, man_Falcon9UsersGuide}. We calculate the wall mass to be 68\% of the whole propellant tank using dimension values from \citet{inPro_Wittal_2022}. As mentioned in an interview \citep{mics_InterviewMusk_engine}, the Merlin 1D engine employs a nickel cobalt inner liner (we assume the CTX alloy series, see above) and most probably a copper alloy jacket \citep{mics_CopperJacket_Falcon9}, for example Narloy-Z or CRCop-84. The nozzle is reported to be made from C-103 niobium alloy.

\subsubsection{Starship upper stage composition}
Due to its enormous dry mass of around 120\,t (see Section \ref{sec:mass_influx})  the Starship upper stage composition is modeled separately. We calculate the mass of the 6 Raptor engines  to make up about 18.0\% of the overall mass, using reported engine dry mass by the manufacturer and assuming a 50\% mass increase for the 3 vacuum version ones due to the nozzle extension. The enormous size of the rocket stage increases the relative mass of the propellant tank/structure, we estimate by 60\%. Starship employs a thermal protection system using ceramic tiles, most probably similar to the Space Shuttle's silica fibre tiles \citep{mics_HRSItile}. From information about the used tile numbers and dimensions \citep{mics_StarshipFlight3}, we infer about 6.7\% of the mass belongs to the thermal protection system, which is reasonably close to estimates in \citet{art_Herberhold_2025}. All this reduces the relative mass of avionics, RCS constrol system, MLI and pressurization tanks compared to the average upper stage. 

The Starship upper stage is constructed from stainless steel; it is highly likely that the composition is close to 304 stainless steel. Furthermore, no paint, insulation foam or payload adapter seems to be used. For all other components, we assume the same materials being used as in the average upper stage, as there is no further detailed information available about the composition of the Starship subsystems. 

\subsubsection{Core stage composition}
While the relative mass of the engines are higher for core stages \citep{art_Schulz_2021}, the components used in both are the same (despite no payload adapter being used in the core stages). We adopt the engine mass of 18\% from \citet{art_Schulz_2021}, and set the propellant tank and structure mass to 60\% due to the larger overall stage size. All other components are renormalized. 

\subsection{Ablation fraction}
We adopt the different ablation rates depending on the type of object from \citet{art_Schulz_2021}, meaning 80\% ablated from satellites and debris, 65\% ablated from all upper stages, and 30\% from core stages. The remaining fractions are expected to impact ground. We do not assume the separate ablation fraction of 100\% for large constellation satellites, because \citet{art_Ott_2025} point out that despite manufacturer claims of complete demise, constellation satellite remnants have been found on ground. While differential ablation is important --- different materials have different melting and vaporization temperatures --- no satisfactory data is available to integrate this into the calculations for the time being.

\subsection{Uncertainties} \label{sec:errors}
The presented data are gathered from a wide range of sources, each with their own uncertainties. In the vast majority of cases, no uncertainties are provided. Furthermore, some values have to be assumed based on sensible approximations and extrapolations, because of the lack of data. This means that for most of the presented values, no uncertainty ranges are readily available. In order to get an impression of the resilience of the injection values, which are calculated from mass influx, composition and ablation fraction, we therefore estimate reasonable uncertainty ranges by comparison to other data or literature. In need of better data, we assume a normal distribution of the estimator for all used values, which allows the resulting uncertainty ranges to be quoted as 3$\sigma$-levels. Thus, Gaussian error propagation can be used to obtain a 3$\sigma$ uncertainty for the elemental injection estimates.

\subsubsection{Mass influx} 
Although the database of \citet{mics_McDowell_rcat_2025}  meticulously documents orbital reentries, dry masses of objects are partially estimated based on available information and can differ from values in other databases such as DISCOS \citep{mics_Discos} and others \citep[e.\,g. see references in][]{art_Barker_2024}. We estimate the uncertainty range by comparing the DISCOS and RCat annual influx estimates. It is important to note that the values from DISCOS are inflated due to satellite wet masses (masses including initial propellant) often being used instead of dry mass estimates. We determine the maximum deviation of both databases by calculating the maximum relative difference after normalizing the DISCOS values to the mean of the RCat data. That way, the wet mass bias is taken out of the equation. The maximum relative difference amounts to 3.5\%, which we conservatively adopt as the uncertainty range. We also adopt this value for core stages.

\subsubsection{Elemental composition} 
While data from numerous different sources could be combined, there are also parts of the estimation where no data could be found and educated assumptions have to be made. We refrain from estimating uncertainties for all the different numbers gathered to come up with a combined, average elemental composition of satellites, different upper stages and core stages, but instead adopt a general uncertainty for different elements, depending on their maturity level. We introduce 4 different maturity levels: for elements which are abundant in many different material groups and components, and have well documented elemental compositions, we estimate a 3$\sigma$ uncertainty equal to 40\% of the elemental mass fraction. We also apply this to elements, which are known to be present in only one or a small number of components, but where we have reason to believe in a good estimate (e.\,g. due to data maturity). Elements H, Li, B, C, N, O, Mg, Al, Si, P, Ti, Cr, Mn, Fe, Ni, Cu, Zn, Mo, and Sn are categorized in this group. 

For medium level maturity, we adopt a 3$\sigma$ uncertainty of 80\% of the elemental mass fraction. Here we group elements for which some credible information is available, but still the mass portion could differ more strongly. Such are Be, F, Na, S, Cl, K, Ca, V, Ga, Ge, As, Br, Zr, Nb, Ag, In, Ba, La, Ce, Hf, Ta, W and Pb. 

For elements with absolute injection masses below 0.05\,t, we only provide an upper limit because an unusually pronounced use of this specific element in only a few spacecraft would result in a large change of the actual injection value.

Lastly, elements where either there is little to no information, the elemental mass portion is 0 or very high uncertainties are expected, are disregarded. This group includes all the other elements and we do not consider those further.

\subsubsection{Ablation fraction} 
Ablation values are still uncertain due to a lack of dedicated in-flight experiments and observations. There exist no detailed measurements of survival portions which would allow  statistical  uncertainties to be evaluated, but estimated values come from findings of remnants of rocket bodies or satellites as well as modeling. The latter  are not validated against measurements and thus the results have to be handled with caution. An overview study by \citet{art_Pardini_2025} assumes a range of 70 to 95\% ablated mass fraction for orbital objects, while \citet{inPro_Ailor_2005} work with 60 to 90\%. Based on this, we conservatively estimate the 3$\sigma$-level of uncertainty at 15\% for upper stages and satellites. For suborbital stages, the uncertainty is higher, as much fewer findings exist. We estimate 20\%, generating a conservative range of 10 to 50\% ablated mass fraction.

\section{Natural injection}
We adapt the estimate of the annual meteoric mass influx to Earth atmosphere by \citet{art_Schulz_2021}, who differentiated between dust particles and larger bolides. For the former, some elements in the composition were not estimated. We complement these by using the elemental abundances of CI-chondrites \citep{art_Lodders_2025}, and renormalize to 100\% mass.

\section{Results, comparisons and discussion}
By combining the information about mass influx, composition and ablation efficiency we arrive at estimates of the mass annually injected into Earth's atmosphere by ablating space waste, broken down to different elements. In order to put this injection into perspective, we compare to the estimated annual natural injection of matter from ablating meteoroids as well as measurements of elements in stratospheric aerosols. 

\subsection{Total influx and injection}
\begin{figure}[!h]
\centering
\includegraphics[scale=0.4]{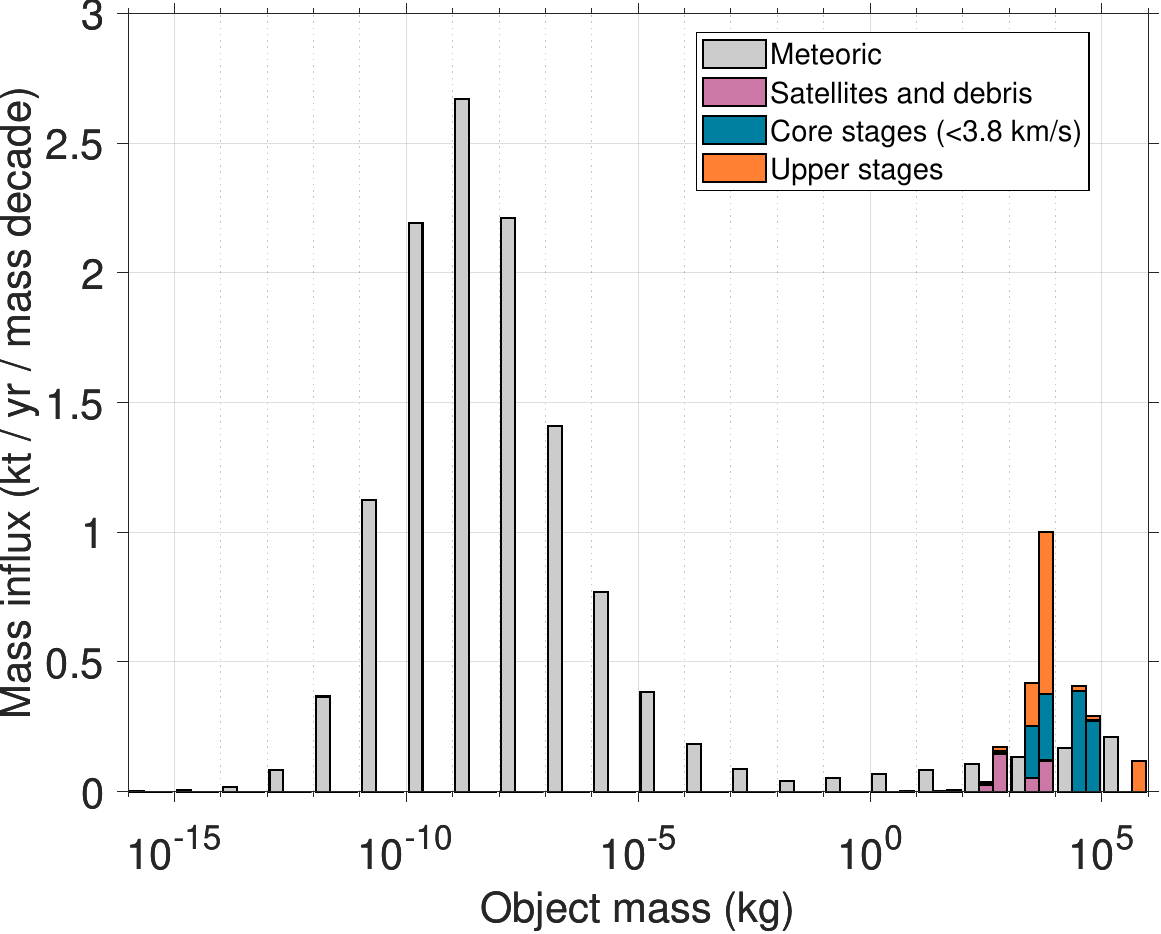}
\caption{Variation of the annual mass influx to the top of the atmosphere with the object mass. The object mass is binned into mass decades. Three different mass influxes are shown: The gray bars depict the annual meteoric mass influx using values from  \citep{art_Schulz_2021}, which is the mean model by \citet{art_Drolshagen_2017}. The middle bar shows the space waste mass influx in 2020, the right bar for 2024, both broken down by object type. Clearly, the space waste influx consists of large objects compared to the meteoric influx.}
\label{fig:massflux_comp}
\end{figure}

As shown in Figure \ref{fig:massflux_comp}, the space waste mass influx (colored bars showing years 2020 and 2024) to the top of the atmosphere is still considerably smaller than the the annual meteoric influx (gray bars). Clearly, there has been as strong increase of space waste mass influx in only the past 4 years due to increased launch activity (especially because of large satellite constellations), that led to more upper stages reentering and also the onset of large constellation satellites reentering after their end of life. Clearly, space waste mass influx can be differentiated quite well by its size, satellites mostly being 0.1 to 10\,t in mass, core stages 1--100\,t, and upper stages mostly in the 1--10\,t range. With the launch trials of Starship, from 2023 onward, objects with masses above 100\,t have started to add to destructive reentry. These large masses are very different from the  meteoric input. The latter dominates the mass influx with trillions of dust particles with masses of the order of $\mu$g. 

\begin{figure}[!h]
\centering
\includegraphics[scale=0.65]{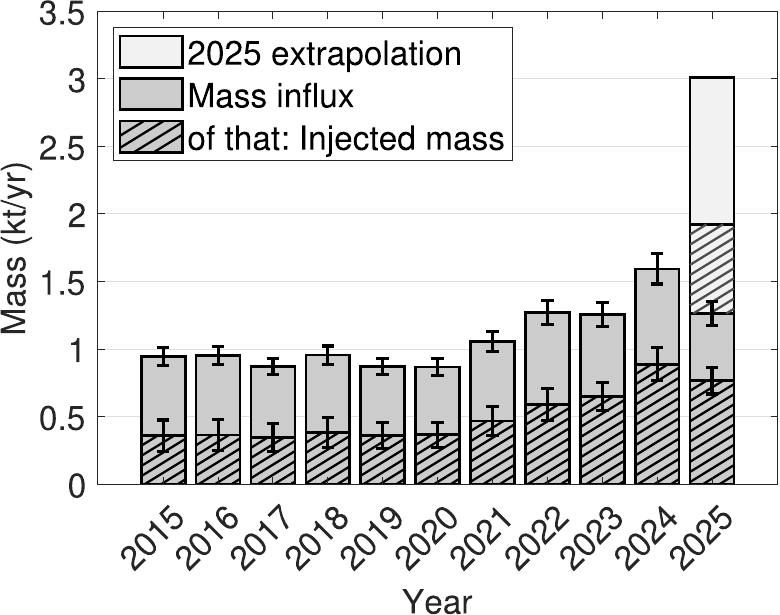}
\caption{Annual space waste mass influx to the top of the atmosphere and the fraction of that, which ablated in the atmosphere (hatched area). For 2025, an extrapolation for the rest of the year (after July 15th 2025) is shown in the lighter gray, again with the hatched area representing the ablated mass. All mass not ablated reaches Earth's surface.}
\label{fig:Inj_Mass}
\end{figure}

The space waste mass influx to the top of the atmosphere, together with the fraction, that ablates and is injected into the atmosphere are plotted in Figure \ref{fig:Inj_Mass}. The corresponding numbers are given in Tables \ref{tab:mass_influx} for the mass influx and \ref{tab:elementgroup} for the injected mass (values in parenthesis are the percentage compared to the natural injection). While the mass influx has been close to 1\,kt in the years 2015 to 2020 (see  Table \ref{tab:mass_influx}), the annual injected mass was only at 340 to 380\,t/yr. In general, the injected mass makes up 40 to 60\% of the mass influx, depending on the relative amount of satellites and the different rocket stages. The rest survives reentry and impacts the ground. From 2021 onward, an increase in mass influx as well as injected mass can be observed. In 2024, 887 $\pm$ 123\,t of material was injected into the atmosphere. Extrapolating from the current data, for 2025 we expect around 1.4\,kt, very close to the probable scenario (Scenario 1) of \citet{art_Schulz_2021}. In the maximum scenario (Scenario 2) of \citet{art_Schulz_2021}, an annual injected mass of 4.4\,kt is to be expected, about a third of the annual natural inejcted mass. 

\subsection{Injection by elements}
\begin{table*}[ht]
  \centering
  \caption{Injected mass (meaning post-ablation) by element group in t. Percentages in parentheses are relative to the meteoric input. The uncertainty for the total injected mass is calculated according to Section \ref{sec:errors}. \newline *until 16th of July 2025. \newline **extrapolated to end of 2025.}
  \label{tab:elementgroup}
  \begin{tabular}{l|rr|rr|rr|rrr}
    \hline
    Year/ \\Scenario & \multicolumn{2}{c|}{Metals (t)}    & \multicolumn{2}{c|}{Metalloids (t)} & \multicolumn{2}{c|}{Non‐metals (t)} & \multicolumn{3}{c}{Total (t)}       \\
    \hline
    2015    &  283& (6)  &   1 &(<1)   &   75 &(1)   &  359& $\pm$ 116  &(3)  \\
    2016    &  289& (7)  &   1 &(<1)   &   74 &(1)   &  364& $\pm$ 115 &(3)  \\
    2017    &  272& (6)  &   1 &(<1)   &   71 &(1)   &  344& $\pm$ 102 &(3)  \\
    2018    &  299& (7)  &   1 &(<1)   &   81 &(1)   &  382& $\pm$ 111 &(3)  \\
    2019    &  282& (6)  &   1 &(<1)   &   79 &(1)   &  363& $\pm$ \enspace96  &(3)  \\
    2020    &  288& (7)  &   1 &(<1)   &   76 &(1)   &  366& $\pm$ \enspace93  &(3)  \\
    2021    &  364& (8)  &   1 &(<1)   &  102 &(2)   &  467& $\pm$ 107 &(4)  \\
    2022    &  456& (10) &   2 &(<1)   &  131 &(2)   &  589& $\pm$ 121 &(5)  \\
    2023    &  515& (12) &   2 &(<1)   &  130 &(2)   &  647& $\pm$ 105 &(5)  \\
    2024    &  688& (16) &   3 &(<1)   &  196 &(3)   &  887& $\pm$ 123 &(7)  \\
    \textit{2025}*    &  \textit{611} & \textit{(14)} &   \textit{4} & \textit{(<1)}   &  \textit{152} & \textit{(2)}   & \textit{766} & \textit{$\pm$ \enspace98} &\textit{(6)}  \\
    \textit{2025}**    & &  &      &   &   & &  \textit{1426} & & \textit{(12)}  \\
    Scen.\,1 & 1025 &(23) &   6 &(<1)  &  427 &(7)   & 1458 & $\pm$ 239 &(12) \\
    Scen.\,2 & 3079 &(70) &  19 &(1)   & 1328 &(22)  & 4426 & $\pm$ 696 &(36) \\
    \hline
    Natural & 4369 & & 1788 & & 6167 & &12325 & \\
    \hline
  \end{tabular}
\end{table*}

While the total injected mass from space waste is an order of magnitude below the natural injection even during 2024 (7\%), the relative comparison looking at the different elements groups reveals that it is more than double that value for metals (14\% compared with the natural input). This is due to the higher portion of metals in space waste and the considerably lower melting temperatures of metals compared with the metal silicates of which meteoroids are mostly composed. This is also the case for other years and both scenarios, as shown in Table \ref{tab:elementgroup}. Metalloids (B, Si, Ge, As, Sb and Te) are dominated by the natural input due to the high amount of silicon in meteoroids; this is also the case with non-metals, which are mostly organic elements highly abundant in meteoroids.

\begin{table*}[!t]
\centering
\caption{Annual elemental mass injection (meaning post-ablation) estimates for the average of 2015–2020, the years 2021 to 2024 and the two future scenarios together with the annual meteoric injection in t. Space waste estimates 100–200\% of the meteoric flux are marked yellow, $\geq$200\% orange, and $\geq$500\% red. The uncertainty is the 3$\sigma$ estimate.}
\label{tab:InjElements}
\small
\begin{tabular}{llllllll|l}
\hline
El.\ 
 & Mean 2015–2020 (t)
 & 2021 (t)
 & 2022 (t)
 & 2023 (t)
 & 2024 (t)
 & Scen.\,1 (t) 
 & Scen.\,2 (t)
 & Meteoric (t) \\ 
\hline
H   & $ 3.0\pm1.1$      & $ 4.0\pm1.2$      & $ 5.2\pm1.5$      & $ 5.1\pm1.3$      & $ 7.8\pm2.0$      & $16.6\pm5.7$     & $51.7\pm18.2$    & 221  \\
Li  & \cellcolor{red!65}$ 0.5\pm0.2$  & \cellcolor{red!65}$ 0.7\pm0.2$  & \cellcolor{red!65}$ 1.0\pm0.3$  & \cellcolor{red!65}$ 1.1\pm0.3$  & \cellcolor{red!65}$ 1.7\pm0.5$  & \cellcolor{red!65}$ 2.7\pm1.0$  & \cellcolor{red!65}$ 8.4\pm3.2$  & 0.02    \\
Be  & \cellcolor{red!65}$<0.05$ & \cellcolor{red!65}$<0.05$ & \cellcolor{red!65}$<0.05$ & \cellcolor{red!65}$<0.05$ & \cellcolor{red!65}$<0.05$ & \cellcolor{red!65}$<0.05$ & \cellcolor{red!65}$ 0.2\pm0.1$  & $3\times10^{-4}$    \\
B   & \cellcolor{red!65}$ 0.1\pm0.0$  & \cellcolor{red!65}$ 0.2\pm0.1$  & \cellcolor{red!65}$ 0.2\pm0.1$  & \cellcolor{red!65}$ 0.2\pm0.1$  & \cellcolor{red!65}$ 0.4\pm0.1$  & \cellcolor{red!65}$ 1.0\pm0.4$  & \cellcolor{red!65}$ 3.1\pm1.2$  & 0.009    \\
C   & $51.2\pm19.8$    & $67.9\pm21.9$    & $86.7\pm26.1$    & $84.5\pm22.0$    & $127.0\pm32.4$   & $276.5\pm94.2$   & $857.0\pm297.8$  & 1154 \\
N   & $ 2.8\pm1.2$     & $ 3.6\pm1.3$     & $ 4.6\pm1.5$     & $ 4.6\pm1.2$     & $ 6.6\pm1.6$     & $12.2\pm4.0$     & \cellcolor{yellow}$37.1\pm12.4$ & 23   \\
O   & $15.8\pm5.7$     & $21.4\pm6.5$     & $27.9\pm8.0$     & $30.1\pm7.1$     & $44.9\pm11.0$    & $94.4\pm33.3$    & $294.9\pm106.1$  & 4169 \\
F   & \cellcolor{orange}$ 2.5\pm1.4$ & \cellcolor{orange}$ 3.4\pm1.9$ & \cellcolor{orange}$ 4.7\pm2.7$ & \cellcolor{orange}$ 4.4\pm2.3$ & \cellcolor{red!65}$ 7.5\pm4.4$  & \cellcolor{red!65}$21.1\pm15.6$ & \cellcolor{red!65}$67.2\pm50.4$  & 1    \\
Na  & $<0.05$     & $0.1\pm0.0$     & $0.1\pm0.1$     & $ 0.1\pm0.1$     & $ 0.2\pm0.1$     & $ 0.5\pm0.4$     & $ 1.6\pm1.3$     & 64   \\
Mg  & $ 1.7\pm0.6$     & $ 2.3\pm0.7$     & $ 3.1\pm0.9$     & $ 2.9\pm0.8$     & $ 4.7\pm1.3$     & $11.5\pm4.3$     & $36.3\pm13.8$    & 1404 \\
Al  & \cellcolor{yellow}$175.7\pm83.6$ & \cellcolor{yellow}$226.5\pm86.8$ & \cellcolor{orange}$285.4\pm99.4$ & \cellcolor{orange}$287.1\pm85.4$ & \cellcolor{orange}$397.0\pm106.1$ & \cellcolor{orange}$618.6\pm212.8$ & \cellcolor{red!65}$1855.5\pm631.4$ & 142  \\
Si  & $ 6.2\pm2.4$     & $ 8.1\pm2.7$     & $10.5\pm3.2$     & $13.1\pm3.1$     & $18.2\pm4.2$     & $32.3\pm11.1$    & $100.1\pm35.0$   & 1788 \\
P   & $ 0.3\pm0.1$     & $ 0.5\pm0.1$     & $ 0.6\pm0.2$     & $ 0.6\pm0.2$     & $ 1.0\pm0.3$     & $ 3.0\pm1.2$     & $ 9.6\pm3.9$     & 23   \\
S   & $<0.05$     & $<0.05$     & $0.1\pm0.0$     & $0.1\pm0.0$     & $ 0.1\pm0.0$     & $ 0.2\pm0.1$     & $ 0.5\pm0.3$     & 559  \\
Cl  & $ 0.1\pm0.1$     & $ 0.2\pm0.1$     & $ 0.2\pm0.1$     & $ 0.2\pm0.1$     & $ 0.3\pm0.1$     & $ 0.7\pm0.4$     & $ 2.2\pm1.4$     & 15   \\
K   & $<0.05$     & $0.1\pm0.0$     & $0.1\pm0.1$     & $0.1\pm0.1$     & $ 0.1\pm0.1$     & $ 0.5\pm0.4$     & $ 1.5\pm1.2$     & 8    \\
Ca  & $ 0.4\pm0.2$     & $ 0.5\pm0.2$     & $ 0.7\pm0.3$     & $ 0.7\pm0.3$     & $ 1.0\pm0.4$     & $ 2.1\pm1.3$     & $ 6.5\pm4.2$     & 94   \\
Ti  & \cellcolor{yellow}$14.2\pm6.1$  & \cellcolor{orange}$18.7\pm6.5$  & \cellcolor{orange}$23.7\pm7.5$  & \cellcolor{orange}$25.6\pm6.6$  & \cellcolor{orange}$35.8\pm8.6$  & \cellcolor{red!65}$60.2\pm19.8$  & \cellcolor{red!65}$183.5\pm60.7$ & 8    \\
V   & $ 0.6\pm0.4$     & $ 0.8\pm0.4$     & \cellcolor{yellow}$1.1\pm0.5$   & \cellcolor{yellow}$1.1\pm0.5$   & \cellcolor{yellow}$1.5\pm0.6$   & \cellcolor{orange}$2.5\pm1.3$   & \cellcolor{red!65}$7.4\pm4.1$    & 1    \\
Cr  & $ 7.5\pm4.4$     & $ 8.8\pm4.4$     & $10.2\pm4.8$     & $19.4\pm6.0$     & $21.9\pm6.2$     & $20.5\pm9.0$     & \cellcolor{yellow}$59.4\pm25.6$  & 40   \\
Mn  & $ 0.8\pm0.4$     & $ 1.0\pm0.4$     & $ 1.3\pm0.5$     & $ 1.7\pm0.5$     & $ 2.2\pm0.5$     & $ 2.5\pm0.9$     & $ 7.3\pm2.7$     & 23   \\
Fe  & $28.4\pm16.1$    & $33.9\pm16.2$    & $40.0\pm18.0$    & $72.3\pm21.6$    & $83.2\pm22.6$    & $82.4\pm34.1$    & $240.8\pm97.3$   & 2477 \\
Co  & $ 2.8\pm1.7$     & $ 3.6\pm1.9$     & $ 4.7\pm2.3$     & \cellcolor{yellow}$5.2\pm2.2$   & \cellcolor{yellow}$7.3\pm3.0$   & \cellcolor{orange}$11.3\pm6.4$  & \cellcolor{red!65}$34.4\pm20.0$   & 5    \\
Ni  & $11.8\pm6.7$     & $14.2\pm6.7$     & $17.2\pm7.5$     & $24.0\pm6.9$     & $29.0\pm7.5$     & $33.8\pm14.2$    & \cellcolor{yellow}$99.2\pm40.7$  & 96   \\
Cu  & \cellcolor{red!65}$23.7\pm11.3$ & \cellcolor{red!65}$29.8\pm11.6$ & \cellcolor{red!65}$37.0\pm13.2$ & \cellcolor{red!65}$37.6\pm10.7$ & \cellcolor{red!65}$52.1\pm13.3$ & \cellcolor{red!65}$97.7\pm33.3$  & \cellcolor{red!65}$297.2\pm101.2$& 2    \\
Zn  & $ 3.1\pm1.3$     & $ 4.1\pm1.4$     & \cellcolor{yellow}$5.2\pm1.6$   & \cellcolor{yellow}$5.1\pm1.4$   & \cellcolor{yellow}$7.3\pm1.9$   & \cellcolor{orange}$13.7\pm4.5$  & \cellcolor{red!65}$41.9\pm13.7$   & 5    \\
Ga  & $<0.05$     & $<0.05$     & $<0.05$     & $<0.05$     & $<0.05$     & $ 0.1\pm0.1$     & \cellcolor{yellow}$0.4\pm0.3$    & 0.2    \\
Ge  & $ 0.3\pm0.2$     & $ 0.4\pm0.4$     & \cellcolor{yellow}$0.6\pm0.5$   & \cellcolor{yellow}$ 0.5\pm0.4$     & \cellcolor{orange}$1.0\pm0.9$   & \cellcolor{red!65}$3.8\pm3.1$    & \cellcolor{red!65}$12.3\pm10.1$   & 0.5    \\
As  & $<0.05$     & $<0.05$     & $<0.05$     & $<0.05$     & $<0.05$     & $ 0.1\pm0.1$     & \cellcolor{yellow}$0.3\pm0.2$    & 0.2    \\
Br  & $ 0.4\pm0.2$     & $ 0.6\pm0.3$     & $ 0.7\pm0.3$     & $ 0.8\pm0.3$     & \cellcolor{yellow}$1.1\pm0.5$   & \cellcolor{orange}$2.2\pm1.4$   & \cellcolor{red!65}$6.8\pm4.3$     & 1    \\
Zr  & \cellcolor{yellow}$0.2\pm0.1$   & \cellcolor{yellow}$0.2\pm0.1$   & \cellcolor{yellow}$0.3\pm0.2$   & \cellcolor{yellow}$0.3\pm0.2$   & \cellcolor{orange}$0.4\pm0.2$   & \cellcolor{orange}$0.6\pm0.3$   & \cellcolor{red!65}$1.8\pm1.0$     & 0.2    \\
Nb  & \cellcolor{red!65}$3.4\pm2.4$    & \cellcolor{red!65}$4.8\pm2.8$    & \cellcolor{red!65}$6.7\pm3.9$    & \cellcolor{red!65}$8.9\pm5.1$    & \cellcolor{red!65}$11.5\pm6.9$   & \cellcolor{red!65}$5.7\pm4.7$  & \cellcolor{red!65}$16.1\pm13.1$   & 0.004    \\
Mo  & \cellcolor{red!65}$0.5\pm0.3$    & \cellcolor{red!65}$0.6\pm0.3$    & \cellcolor{red!65}$0.7\pm0.4$    & \cellcolor{red!65}$0.8\pm0.3$    & \cellcolor{red!65}$1.0\pm0.3$    & \cellcolor{red!65}$1.3\pm0.7$    & \cellcolor{red!65}$3.8\pm1.9$     & 0.01    \\
Ag  & \cellcolor{red!65}$0.7\pm0.5$    & \cellcolor{red!65}$1.0\pm0.6$    & \cellcolor{red!65}$1.5\pm0.9$    & \cellcolor{red!65}$1.4\pm0.8$    & \cellcolor{red!65}$2.4\pm1.5$    & \cellcolor{red!65}$6.9\pm5.3$    & \cellcolor{red!65}$22.2\pm17.2$   & 0.003   \\
In  & \cellcolor{orange}$<0.05$     & \cellcolor{orange}$<0.05$     & \cellcolor{red!65}$<0.05$     & \cellcolor{red!65}$<0.05$     & \cellcolor{red!65}$<0.05$     & \cellcolor{red!65}$<0.05$     & \cellcolor{red!65}$0.1\pm0.1$     & $9\times10^{-4}$\\
Sn  & \cellcolor{red!65}$1.8\pm0.7$    & \cellcolor{red!65}$2.5\pm0.8$    & \cellcolor{red!65}$3.2\pm0.9$    & \cellcolor{red!65}$3.4\pm0.8$    & \cellcolor{red!65}$5.0\pm1.2$    & \cellcolor{red!65}$9.7\pm3.3$    & \cellcolor{red!65}$30.2\pm10.5$    & 0.02    \\
Ba  & \cellcolor{orange}$0.1\pm0.1$    & \cellcolor{orange}$0.1\pm0.1$    & \cellcolor{red!65}$0.2\pm0.1$    & \cellcolor{red!65}$0.2\pm0.1$    & \cellcolor{red!65}$0.3\pm0.2$    & \cellcolor{red!65}$0.8\pm0.6$    & \cellcolor{red!65}$2.4\pm1.8$     & 0.04    \\
La  & $<0.05$     & $<0.05$     & $<0.05$     & $<0.05$     & $<0.05$     & $<0.05$     & \cellcolor{orange}$<0.05$     & 0.003   \\
Ce  & $<0.05$     & \cellcolor{yellow}$<0.05$     & \cellcolor{orange}$<0.05$     & \cellcolor{yellow}$<0.05$     & \cellcolor{orange}$<0.05$     & \cellcolor{red!65}$0.1\pm0.1$    & \cellcolor{red!65}$0.3\pm0.3$     & 0.009   \\
Hf  & \cellcolor{red!65}$0.3\pm0.2$    & \cellcolor{red!65}$0.5\pm0.3$    & \cellcolor{red!65}$0.7\pm0.4$    & \cellcolor{red!65}$0.9\pm0.5$    & \cellcolor{red!65}$1.2\pm0.7$    & \cellcolor{red!65}$0.5\pm0.4$    & \cellcolor{red!65}$1.5\pm1.2$     & 0.002   \\
Ta  & \cellcolor{red!65}$<0.05$     & \cellcolor{red!65}$0.1\pm0.0$    & \cellcolor{red!65}$0.1\pm0.0$    & \cellcolor{red!65}$0.1\pm0.0$     & \cellcolor{red!65}$0.1\pm0.1$    & \cellcolor{red!65}$0.2\pm0.1$    & \cellcolor{red!65}$0.6\pm0.3$     & $2\times10^{-4}$\\
W   & \cellcolor{red!65}$0.5\pm0.4$    & \cellcolor{red!65}$0.6\pm0.4$    & \cellcolor{red!65}$0.7\pm0.4$    & \cellcolor{red!65}$0.9\pm0.4$    & \cellcolor{red!65}$1.1\pm0.5$    & \cellcolor{red!65}$1.1\pm0.8$    & \cellcolor{red!65}$3.0\pm2.1$     & 0.001   \\
Pb  & \cellcolor{red!65}$ 1.2\pm0.6$     & \cellcolor{red!65}$ 1.6\pm0.8$     & \cellcolor{red!65}$ 2.1\pm1.0$     & \cellcolor{red!65}$ 2.2\pm0.9$     & \cellcolor{red!65}$3.2\pm1.4$   & \cellcolor{red!65}$6.1\pm3.7$    & \cellcolor{red!65}$19.0\pm11.8$   & 0.03    \\
\hline
\end{tabular}
\end{table*}

Table \ref{tab:InjElements} shows the injected mass differentiated by elements. Uncertainties have been calculated by applying Gaussian error propagation on the  uncertainty estimates of Section \ref{sec:errors}. A 3$\sigma$ level of certainty is given to reflect an interval within which injection values are lying with very high probability. As uncertainties were not determined statistically but had to be estimated, it is possible that actual values can still lie outside this interval. The annual injection masses for an average of 2015 to 2020 and each year from 2021 up to 2024, as well as the two scenarios, are compared to the natural injection: values in yellow show injection masses surpassing the estimated annual meteoric input up to a factor of 2; orange depicts a factor of 2--5 times higher; and red at least 5 times the meteoric injected mass. Clearly, the abundant use of certain metal elements leads to a large number of them dominating the natural input. In 2015, 18 metal elements dominated the natural injection, while in 2024 this increased to 24 elements. In a 75,000 large constellation satellite scenario (Scenario 2), this number could increase to 30. The notable element with a high absolute anthropogenic input is Al, but also C, Fe, Cu, O, Ti, Ni, Cr and even Nb and Si. Of these, C, Fe, O and Si have a much larger natural input due to their high abundance in meteoroids; the anthropogenic injection is currently not relevant. Ni and Cr are only comparable to the meteoric injection for Scenario 2. The space waste injection of Al, Cu, Ti and Nb, however, has dominated the meteoric injection since at least 2015. In particular, the transition metals Cu, Ti and Nb are potentially capable of introducing new chemistry pathways in Earth's atmosphere, because these transition metals can have at least two accessible oxidation states and hence participate in catalytic cycles. While a number of the anthropogenically dominating elements have injection values below 10\,t/yr --- orders of magnitude lower than tropospheric emissions \citep[see e.\,g.][]{art_Zhu_2020} --- they are much less abundant in the meteoric input. Note that although the total mass input is low, the amounts are sufficient to be detected in the atmosphere and thus these elements can serve as tracers for space waste pollution of the atmosphere. Relevant elements are Li, B, F, V, Co, Zn, Ge, Br, Zr, Mo, Ag, Sn, Ba, Hf, Ta, W and Pb. While Be, In and Ce dominate over the natural influx as well, the general amount of these elements in meteoroids as well as anthropogenic materials is extremely low; the anthropogenic injection is estimated below 50\,kg in 2024. 

Comparison of the presented injection values with previous estimates for 2019 by \citet{art_Schulz_2021} shows them strongly underestimating carbon and magnesium injection. Here, several organic-containing materials like adhesives, foams or CFRP have been considered; magnesium can be found in optical materials and several alloys not considered in the previous study. In contrast, \citet{art_Schulz_2021} overestimated the use of germanium by an order of magnitude --- this may be attributed to a substantial overestimation of the solar cell mass of a typical satellite. The previous estimates of Al, Si, Cr, Fe, and Cu are within the error ranges presented here, while titanium injection is estimated half and nickel injection double of the present estimate. This may be attributed to a much more detailed breakdown of used alloys in the present study. Overall, the present estimates constitute a substantial improvement over the previous estimates. In the following, specific elements will be addressed individually.

\paragraph{Lithium}  
Al-Li alloys and Li-Ion batteries are the main contributors. Its abundance especially in rocket bodies might increase further in the coming years and decades, as Al-Li alloys are used more frequently. Lithium is especially interesting because it can be detected in the upper mesosphere above 80 km by lidar \citep{art_Hauchecorne_1980, art_Jegou_1980} , and measurements were made well before anthropogenic injection levels were significant, because neither Al-Li alloys nor Li-ion batteries being used in the space industry. Therefore, current measurements allow for a comparison to natural levels. 

\paragraph{Aluminum} 
Aluminum is the main element in space waste. Due to its light mass, it is the preferred structural material in raw or alloy form. Therefore, it already dominates the meteoric injection since at least 2015. Due to its high mass input compared to other elements, it has been the focus of past studies \citep{art_Schulz_2021, art_Shutler_2022, art_Miraux_2022, art_Barker_2024, art_Ferreira_2024}. Possible effects on the atmosphere include changed radiative forcing from aerosol particles, but also direct and catalytic destruction of ozone on alumina (Al$_2$O$_3$) particle surfaces \citep{art_Lee_1996, art_Molina_1997}. While the dominance of alumina from space waste ablation is proposed by equilibrium chemistry modeling post-ablation \citep{art_Park_2021}, it is questionable whether this is oversimplified. Meteoroid ablation and atmospheric chemistry modeling suggest AlOH and Al(OH)$_2$ as the dominating chemical species at altitudes below 80\,km \citep{art_Plane_2021}, which is supported by the preliminary analysis of lab ablation experiments \citep{misc_GomezMartin_2025}. The hydroxide species in particular suggest completely different pathways of influencing ozone levels as well as other atmospheric effects, yet to be determined.

\paragraph{Fluorine} 
Fluorine is present in rubbers and plastics, but especially in the electrolyte of lithium ion batteries as lithium hexafluorophosphate (LiPF6). It has been detected often in stratospheric aerosol particles with space waste metals \citep{art_Murphy_2023}, and its dominance over the meteoric input raises questions about possible atmospheric chemistry. A possible pathway could be the synthesis of hydrogen fluoride. However, the injected amount of fluorine is very low compared to the amount of fluorine (the majority from CFCs) already in the atmosphere \citep{art_Raymond_2025} --- about 1 ppb translates to roughly $10^6$\,t in the stratosphere, a factor of $10^5$ more than the space waste injection in 2024. For the mesosphere, due to its lower mass, this ratio drops to 500. In the thermosphere, the total fluorine is comparable to the total injection today. However, most of space waste is ablated below the thermosphere and is subsequently transported downwards.

\paragraph{Chlorine and bromine} 
The only sources of chlorine determined in this study include remnants in epoxy resins which are used in CFRP and as adhesives, coatings etc. We did not find that they are usually part of the newer cyanate ester resins, contrary to  \citet{techrep_ATISPADE_2019}. Standard plastics like polychlorotrifluoroethylene (PCTFE) or polyvinyl chloride (PVC) naturally have chlorine in them, but seem to not be used regularly in space applications due to strong outgassing \citep{mics_spacematdb}. As a result, the total chlorine injection by space waste is far lower than the meteoric input, even in extreme scenarios. This points towards the direct impact on for example ozone levels being too small to be measurable. 

Bromine on the other hand, being present in PCBs, began to dominate over the naturally injected bromine in 2024 and might exceed the natural injection by a factor of 5 in the future. However, assuming a residence time of 5 years, the mass input of less than 5\,t is very low compared to the bromine currently present in different atmospheric layers. Using a bromine concentration of 20 ppt \citep{art_Wales_2021} we arrive at a total bromine mass of about 15\,kt in the stratosphere.

\paragraph{Arsenic}
Contrary to presentations of preliminary results \citep{misc_Schulz_2024, misc_Schulz_2025} which were based on an inflated estimate of arsenic content in solar cell material \citep{techrep_ATISPADE_2019}, the arsenic injection does not dominate over the natural injection. The compositional refinement using data sheets, and the literature of triple-junction solar cell materials, indicates that only a small amount of arsenic is being used. Thus, only in the extreme scenario do space waste injection levels dominate over the natural input from meteoroids. Compared to the annual tropospheric release of about 31\,kt per year \citep{art_Walsh_1979}, the anthropogenic atmospheric injection of less than 50\,kg is negligible when it comes to toxicity concerns.

\paragraph{Transition metals} 
Transition metals are known for their catalytic activity, for example as as organic derivative or hydrate and oxide catalysts. The domination over the natural input of Ti, Co, Cu, Zn, Zr, Nb, Mo, Ag, Hf, Ta, W shows the need for a better understanding of the chemical pathways of these metals, especially with regards to cloud nucleation and stratospheric heterogeneous chemistry.

\subsection{Comparison to stratospheric data}
\begin{table}[h]
\centering
\caption{Element injection (meaning post-ablation) relative to aluminum injection in 2022 for selected elements and the comparison values from \citet{art_Murphy_2023}. Values are given in \%.}
\label{tab:relAlComp}
\begin{tabular}{ldldl}
\hline
Element & \multicolumn{2}{l}{\begin{tabular}[c]{@{}l@{}}Relative ratio to \\ Al (this study)\end{tabular}} & \multicolumn{2}{l}{\begin{tabular}[c]{@{}l@{}}Relative ratio to Al \\ (Murphy et al. 2023)\end{tabular}} \\
\hline
Li  & 0.4                                     & $\pm$ 0.2                                     & 0.4                                          & $\pm$ 0.2                                          \\
Be  & 0.003                                    & $\pm$ 0.003                                    &                                              &                                        \\
Ti  & 8.3                                      & $\pm$ 3.9                                   &                                              &                                        \\
Cr  & 3.6                                      & $\pm$ 2.1                                  &                                              &                                         \\
Cu  & 13                                     & $\pm$ 7                                     & 12                                           & $\pm$ 6                                            \\
Nb  & 2.3                                      & $\pm$ 1.6                                    &                                              &                                      \\
Mo  & 0.3                                    & $\pm$ 0.2                                    &                                              &    
\\
Ag  & 0.5                                      & $\pm$ 0.4                                  &                                              &                                         \\
Sn  & 1.1                                      & $\pm$ 0.5                                     &                                              &                                        \\
Hf  & 0.2                                     & $\pm$ 0.2                                &                                              &                                          \\
Ta  & 0.03                                     & $\pm$ 0.02                                   &                                              &                                          \\
W   & 0.3                                     & $\pm$ 0.2                                     &                                              &                                         \\
Pb  & 0.7                                      & $\pm$ 0.4                                     & 0.9                                          & $\pm$ 0.5        \\
\hline
\end{tabular}
\end{table}

Due to atmospheric circulation, material from the mesosphere injection altitudes is transported towards the winter pole, where it descends within the polar vortex to the lower stratosphere. \citet{art_Murphy_2023} analyzed individual stratospheric aerosol particles, calculating metal ratios relative to aluminum for the material coming from space waste reentry, excluding the meteoric injection. Thus, we can directly compare the elemental injection rates presented here to their findings. We use the data from 2022 to coincide with their measurements at the beginning of 2023. The comparison is shown in Table \ref{tab:relAlComp}. For all elements, we see excellent agreement; values are well within the limits of uncertainty. This strongly support the findings by \citet{art_Murphy_2023} of space waste remnants being present in stratospheric aerosol particles. By reducing the uncertainties of both methods, insights about possible differential ablation of spacecraft components and thus of different elements can be gained. 

Comparisons to meteoric injection were presented by \citet{art_Murphy_2023} as well. They found more than 70\% of the cumulative aluminum in particles strongly contaminated with space waste elements, and more than 90\% for Li, Cu and Pb. For 2022, our results indicate that the fraction of space waste aluminum in the total injection from all natural and man-made materials is 67\%, 98\% for Li, 94\% for Cu and 99\% for Pb. This excellent agreement points towards a high percentage of space waste ablated material being vaporized  \citep[as hypothesized by][]{art_Murphy_2023}, instead of ablating as molten droplets and ending up in large particulates. Particles of sizes above ($>10\mu$m) sediment to the ground within weeks and would not be detectable in the stratosphere \citep{art_Glassmeier_2025}. On the other hand, particles of sizes between 10 and 100\,nm are transported from mesospheric injection altitudes to the stratosphere  by the residual circulation and can reside there for a few years and thus accumulate \citep{art_Maloney_2025}. By better constraining the uncertainties in both the present study and \citet{art_Murphy_2023}, and better understanding stratospheric particle accumulation, one could possibly work to get closer to a size distribution of space waste ablation emissions. Due to the different residence times in the atmosphere and different total surface area, particle size distributions are a key element in determining the importance of ablated space waste in the middle atmosphere.

The elements suspected of being space waste remnants in \citet{art_Murphy_2023} are also predicted in our study, namely  Li, Al, Cu, Nb, Ag, Hf, Pb, Be, Mg, Ti, Cr, Fe, Ni, Zr, Ga, Mo, In, Sn, Ba, Ce, Ta and  W. Elements not present in significant amounts here, but detected in their study are Sb and Bi, but also the rarely seen Cd, Pr and Nd. A further refinement of the compositional estimate should also focus on these.

\section{Conclusion and outlook}
These updated estimates of the annual influx and injection of space waste matter into Earth's atmosphere from 2015 up to 2025 confirm the anticipated strong increase in mass influx \citep[e.\,g.][]{art_Schulz_2021}. Between 2020 and 2024, the mass influx and injected mass has more than doubled, whereas it stayed almost constant in the years before. This shows that the onset of the large satellite constellation age is the driving factor of the increased space waste reentry into the atmosphere. The future scenarios adopted from \citet{art_Schulz_2021} are slowly becoming reality with the mass injection extrapolated to the end of 2025 equaling the ``probable'' scenario (Scenario 1). 

Although the annually injected mass is still an order of magnitude smaller than that injected by meteoroids, the high abundance of metals in space waste results in a much higher ratio compared to the natural injection (17\% in 2024). In a scenario of 75,000 large constellation satellites --- which is well within reach of the current proposals and activities --- this value could rise to 70\% in the future. This percentage is lower than the original estimate in \citet{art_Schulz_2021} due to an adjustment of the natural input and a much more detailed compositional evaluation of space waste. More importantly, a large number of elements --- mainly metals --- already dominate the natural injection. The number has increased from 18 elements in the 2015 to 24 in 2024, and could further increase to 30 in the future. They are mainly elements with low to medium natural injection levels. The element with the highest mass input by far is aluminum, which is regularly considered in studies. However, other dominating elements should also be considered, even if the injected mass is low. Notable elements include
Cu, Ti, Nb, Co, Zn, Sn, Pb, Ag, Li, V, Hf, W, Ge, Mo, Zr, B, and Ba.

The wide variety of elements injected by reentering space waste, combined with their dominance over the meteoric injection, underscores the need for comprehensive research on associated atmospheric effects. The fact that the presented injection estimates fit very well to measurements of stratospheric aerosol particles suggests that either a large portion of the injected material is vaporized during ablation or that a substantial amount of the injected material stays in the upper atmosphere for prolonged periods of time. The considerable injection of transition metals, which are known for their catalytic activity, raises question about long-term effects on Earth's atmosphere via yet unknown chemical pathways. This adds to existing worries about effects on ozone chemistry in the stratosphere \citep{art_Schulz_2021, art_Shutler_2022, art_Barker_2024, art_Ferreira_2024}, increased cloud nucleation, or climate effects.

The present study considerably improved compared to the original estimates by \citet{art_Schulz_2021}. The consideration of various databases, industry input, material information, and literature allows for the computation of injection values for 43 elements, compared to 10 previously. Values for all other naturally occurring elements were obtained (up to atomic numbers of uranium excluding Tc, At, Fr, Pm and all noble gases) but not considered due to excessive uncertainties. In general, considerable uncertainties are still involved in all parts of the estimation, namely mass influx, space waste composition, and ablation behavior. In order to improve the estimates in the future, more detailed data is needed, especially from companies involved in supplying components and materials, and building and operating satellites and launchers. Precise data is needed on component materials, component mass fractions, dry masses of rocket stages and the trajectories of launchers. Additionally, much more research has to go into destructive ablation. Probably the largest source of error in this study is the neglect of differential ablation, where relatively volatile elements ablate much more rapidly than refractory species \citep{art_Janches_2009}. Also, the differentiated nature of space waste leads to internal components being shielded from ablation. Both these aspects have been observed in wind tunnel experiments \citep{art_Bonvoisin_2023}, as well as from recovered ground debris, and it is known to strongly influence the relative injection of elements for the meteoric input. However, at the moment, there is insufficient data to model space waste differential ablation. Thus, there is a need for dedicated searches for space waste that survived reentry and impacted ground. Additionally, ground experiments representative of the conditions of atmosphere re-entry need to be carried out to understand how much differential ablation influences the input into the atmosphere. 

Today's large satellite constellations amplify the problem of on-orbit space debris as well as the ground risk from impacts. Beside these well-discussed problems, the injection estimates presented here indicate a substantial risk associated with space waste reentry and possible effects on Earth's atmosphere and thus the human habitat.

\section*{Acknowledgments}
The authors want to thank James Beck, Andreas Hördt, Dennis Michael Jöckel, Viktoria Kutnohorsky, Tobias Lips, Martin Ross, and Martin Spel for stimulating discussion and helpful advice.

For our research we are using information from ESA DISCOS (Database and Information System Characterising Objects in Space), a single-source reference for launch information, object registration details, launch vehicle descriptions, as well as spacecraft information for all trackable, unclassified objects. We acknowledge ESA's efforts to maintain and operate this database with its APIs.

\section*{Appendix A: Rocket stages dry mass and re-entry velocity} \label{AppA}
Table \ref{tab:AppA1} gives all rocket stages with their dry mass and re-entry velocity of rockets active between 2015 and 2025. From that, for reach rocket, the total mass of the core stages (suborbital) which destructively re-enter the atmosphere (with velocities larger than 3.8\,km/s, but not reaching orbital speed) can be derived, given in the last column of Table \ref{tab:AppA1}. Upper stages reaching orbit do not add to this total mass. Booster stages (stage number 0) are assumed to have speeds below 3.8\,km/s; so are payload fairings (only for the Long March 4B and 4C). We also disregard all stages that make a controlled landing. Rockets that were launched, but failed to reach orbit (OS-M1, RS-1, Spectrum, Super Strypi (SPARK), Terran-1) as well as suborbital rockets are not considered. Stage mass data were obtained from \citet{art_Schulz_2021}, the German Aerospace Center (DLR), \citet{art_Barker_2024}, \citet{mics_McDowell_gcat_2025}, and \citet{misc_GuntersSpacePage}. Velocity data were obtained from \citet{art_Schulz_2021}, the German Aerospace Center (DLR), \citet{misc_BautzeScherf_2022}, and calculated using trajectory data from launch videos. For some rockets, no information is available. In these cases, values were assumed (marked yellow in the Table) based on similar launch vehicles.

\begin{table*}[!t]
\caption{Rocket stage masses and re-entry velocities of rockets. The total mass of all suborbital stages of a rocket that destructively re-enter at speeds larger than 3.8\,km/s is given in the last column. Assumed values are marked yellow. Certain rockets have different configurations, where the booster stage or the orbital stage differ in dry mass. In this case, the rockets are summarized and a green field depicts possible different configurations. When no data is available and no assumptions were made, the field is marked red.}
\label{tab:AppA1}
\footnotesize
\begin{tabular}{l|l|llll|llll|l}
\hline 
Rocket                                  &  \multicolumn{5}{c}{Stage dry mass (kg)}                                                                                                                                                                          & \multicolumn{4}{c}{\begin{tabular}[c]{@{}c@{}}Stage re-entry velocity \\ at 100\,km (km/s)\end{tabular}}                                                                                                          & \multicolumn{1}{c}{\begin{tabular}[c]{@{}c@{}}Total re-entry mass \\ with v\textgreater{}3.8\,km/s (kg)\end{tabular}} \\
Stage \#                & 0                                               & 1                                               & 2                                                   & 3                                                  & 4                                               & 1                                                                & 2                                                                & 3                                                                & 4        &                                                                                                                       \\ \hline 
Angara 1.2                              & 0                                               & 10150                                           & 2355                                                & 1000                                               &                                                 & \cellcolor[HTML]{FFEB9C}{\color[HTML]{9C6500} \textless{}3.8}    & \cellcolor[HTML]{FFEB9C}{\color[HTML]{9C6500} \textgreater{}3.8} & orbital                                                          &          & 2355                                                                                                                  \\
Angara (A5, A5 Orion, and A5 Persei)    & 39200                                           & 11000                                           & 4000                                                & \cellcolor[HTML]{C6EFCE}{\color[HTML]{006100} }    &                                                 & 5.0                                                              & 7.6                                                              & orbital                                                          &          & 15000                                                                                                                 \\
Antares 230                             & 0                                               & 19700                                           & 1392                                                &                                                    &                                                 & 3.8                                                              & orbital                                                          &                                                                  &          & 19700                                                                                                                 \\
Ariane 5 (ECA and ES)                   & 31000                                           & 14700                                           & \cellcolor[HTML]{C6EFCE}{\color[HTML]{006100} }     &                                                    &                                                 & 6.9                                                              & orbital                                                          &                                                                  &          & 14700                                                                                                                 \\
Ariane 62                               & 22000                                           & 14700                                           & 3000                                                &                                                    &                                                 & 4.5                                                              & orbital                                                          &                                                                  &          & 14700                                                                                                                 \\
Astra Rocket 3                          & 0                                               & 950                                             & 250                                                 &                                                    &                                                 & \cellcolor[HTML]{FFEB9C}{\color[HTML]{9C6500} \textless{}3.8}    & orbital                                                          &                                                                  &          & 0                                                                                                                     \\
Atlas V 4XY                             & \cellcolor[HTML]{C6EFCE}{\color[HTML]{006100} } & 21054                                           & \cellcolor[HTML]{C6EFCE}{\color[HTML]{006100} }     &                                                    &                                                 & 4.9                                                              & orbital                                                          &                                                                  &          & 21054                                                                                                                 \\
Atlas V 5XY                             & \cellcolor[HTML]{C6EFCE}{\color[HTML]{006100} } & 21351                                           & 2030                                                &                                                    &                                                 & 4.9                                                              & orbital                                                          &                                                                  &          & 21351                                                                                                                 \\
Atlas V N22                             & 10200                                           & 22825                                           & 2030                                                &                                                    &                                                 & 4.9                                                              & orbital                                                          &                                                                  &          & 22825                                                                                                                 \\
Ceres-1                                 & 0                                               & 590                                             & 274                                                 & 86                                                 & 2                                               & \cellcolor[HTML]{FFEB9C}{\color[HTML]{9C6500} \textless{}3.8}    & \cellcolor[HTML]{FFEB9C}{\color[HTML]{9C6500} \textless{}3.8}    & \cellcolor[HTML]{FFEB9C}{\color[HTML]{9C6500} \textgreater{}3.8} & orbital  & 86                                                                                                                    \\
Chollima-1                              & 0                                               & \cellcolor[HTML]{FFC7CE}{\color[HTML]{9C0006} } & \cellcolor[HTML]{FFEB9C}{\color[HTML]{9C6500} 500}  & \cellcolor[HTML]{FFC7CE}{\color[HTML]{9C0006} }    &                                                 & \cellcolor[HTML]{FFEB9C}{\color[HTML]{9C6500} \textless{}3.8}    & \cellcolor[HTML]{FFEB9C}{\color[HTML]{9C6500} \textgreater{}3.8} & orbital                                                          &          & 500                                                                                                                   \\
Delta 4H                                & 54400                                           & 27200                                           & 3420                                                &                                                    &                                                 & 6.0                                                              & orbital                                                          &                                                                  &          & 27200                                                                                                                 \\
Delta 4M+ ((4,2), (5,2), and (5,4))     & \cellcolor[HTML]{C6EFCE}{\color[HTML]{006100} } & 27200                                           & \cellcolor[HTML]{C6EFCE}{\color[HTML]{006100} }     &                                                    &                                                 & 4.9                                                              & orbital                                                          &                                                                  &          & 27200                                                                                                                 \\
Delta II (types 7320, 7420, 7920)       & \cellcolor[HTML]{C6EFCE}{\color[HTML]{006100} } & 5680                                            & 950                                                 &                                                    &                                                 & \cellcolor[HTML]{FFEB9C}{\color[HTML]{9C6500} \textgreater{}3.8} & orbital                                                          &                                                                  &          & 5680                                                                                                                  \\
Dnepr                                   & 0                                               & 15000                                           & 3000                                                & 900                                                &                                                 & \cellcolor[HTML]{FFEB9C}{\color[HTML]{9C6500} \textless{}3.8}    & \cellcolor[HTML]{FFEB9C}{\color[HTML]{9C6500} \textgreater{}3.8} & orbital                                                          &          & 3000                                                                                                                  \\
Electron and Electron Curie             & 0                                               & 950                                             & 250                                                 & 40                                                 &                                                 & \cellcolor[HTML]{FFEB9C}{\color[HTML]{9C6500} \textless{}3.8}    & 7.6                                                              & orbital                                                          &          & 250                                                                                                                   \\
Epsilon-2                               & 0                                               & 8700                                            & 2000                                                & 800                                                &                                                 & 2.1                                                              & 5.7                                                              & orbital                                                          &          & 2000                                                                                                                  \\
Epsilon-2 CLPS                          & 0                                               & 8700                                            & 2000                                                & 800                                                & 155                                             & \cellcolor[HTML]{FFEB9C}{\color[HTML]{9C6500} 2.1}               & \cellcolor[HTML]{FFEB9C}{\color[HTML]{9C6500} 5.7}               & \cellcolor[HTML]{FFEB9C}{\color[HTML]{9C6500} $>5.7$}            & orbital  & 2800                                                                                                                  \\
Falcon 9 (v1.1 and v1.2)                & 0                                               & 21100                                           & 3700                                                &                                                    &                                                 & lands                                                            & orbital                                                          &                                                                  &          & 0                                                                                                                     \\
Falcon Heavy                            & 39500                                           & 21300                                           & 4250                                                &                                                    &                                                 & lands                                                            & orbital                                                          &                                                                  &          & 0                                                                                                                     \\
Firefly Alpha                           & 0                                               & 2895                                            & 909                                                 &                                                    &                                                 & 2.9                                                              & orbital                                                          &                                                                  &          & 0                                                                                                                     \\
GSLV Mk II                              & 24000                                           & 30000                                           & 4800                                                & 2200                                               &                                                 & 4.8                                                              & orbital                                                          &                                                                  &          & 30000                                                                                                                 \\
GSLV Mk III                             & 62300                                           & 10100                                           & 4850                                                &                                                    &                                                 & 4.6                                                              & orbital                                                          &                                                                  &          & 10100                                                                                                                 \\
GYUB - TV2 & 0                                               & \cellcolor[HTML]{FFC7CE}{\color[HTML]{9C0006} } & \cellcolor[HTML]{FFEB9C}{\color[HTML]{9C6500} 500}  & \cellcolor[HTML]{FFC7CE}{\color[HTML]{9C0006} }    &                                                 & \cellcolor[HTML]{FFEB9C}{\color[HTML]{9C6500} \textless{}3.8}    & \cellcolor[HTML]{FFEB9C}{\color[HTML]{9C6500} \textgreater{}3.8} & orbital                                                          &          & 500                                                                                                                   \\
Gravity-1                               & \cellcolor[HTML]{FFC7CE}{\color[HTML]{9C0006} } & \cellcolor[HTML]{FFC7CE}{\color[HTML]{9C0006} } & \cellcolor[HTML]{FFEB9C}{\color[HTML]{9C6500} 2000} & \cellcolor[HTML]{FFC7CE}{\color[HTML]{9C0006} }    &                                                 & \cellcolor[HTML]{FFEB9C}{\color[HTML]{9C6500} \textless{}3.8}    & \cellcolor[HTML]{FFEB9C}{\color[HTML]{9C6500} \textgreater{}3.8} & orbital                                                          &          & 2000                                                                                                                  \\
H-IIA (202 and 204)                     & \cellcolor[HTML]{C6EFCE}{\color[HTML]{006100} } & 12900                                           & 3100                                                &                                                    &                                                 & 5.1                                                              & orbital                                                          &                                                                  &          & 12900                                                                                                                 \\
H-IIB                                   & 42200                                           & 24200                                           & 3400                                                &                                                    &                                                 & 5.5                                                              & orbital                                                          &                                                                  &          & 24200                                                                                                                 \\
H-III 22                                & \cellcolor[HTML]{FFC7CE}{\color[HTML]{9C0006} } & 25000                                           & \cellcolor[HTML]{FFC7CE}{\color[HTML]{9C0006} }     &                                                    &                                                 & 3.6                                                              & orbital                                                          &                                                                  &          & 0                                                                                                                     \\
Jielong-1                               & 0                                               & \cellcolor[HTML]{FFC7CE}{\color[HTML]{9C0006} } & \cellcolor[HTML]{FFC7CE}{\color[HTML]{9C0006} }     &                                                    &                                                 & \cellcolor[HTML]{FFEB9C}{\color[HTML]{9C6500} \textless{}3.8}    & orbital                                                          &                                                                  &          & 0                                                                                                                     \\
Jielong-3                               & 0                                               & 8700                                            & 2000                                                & 800                                                & 155                                             & \cellcolor[HTML]{FFEB9C}{\color[HTML]{9C6500} \textless{}3.8}    & \cellcolor[HTML]{FFEB9C}{\color[HTML]{9C6500} \textgreater{}3.8} & \cellcolor[HTML]{FFEB9C}{\color[HTML]{9C6500} \textgreater{}3.8} & orbital  & 2800                                                                                                                  \\
KAIROS                                  & 0                                               & 1769                                            & 522                                                 & 245                                                &                                                 & \cellcolor[HTML]{FFEB9C}{\color[HTML]{9C6500} \textless{}3.8}    & \cellcolor[HTML]{FFEB9C}{\color[HTML]{9C6500} \textgreater{}3.8} & orbital                                                          &          & 522                                                                                                                   \\
Kaituozhe-2                             & 0                                               & \cellcolor[HTML]{FFC7CE}{\color[HTML]{9C0006} } & \cellcolor[HTML]{FFEB9C}{\color[HTML]{9C6500} 1000} & \cellcolor[HTML]{FFC7CE}{\color[HTML]{9C0006} }    &                                                 & \cellcolor[HTML]{FFEB9C}{\color[HTML]{9C6500} \textless{}3.8}    & \cellcolor[HTML]{FFEB9C}{\color[HTML]{9C6500} \textgreater{}3.8} & orbital                                                          &          & 1000                                                                                                                  \\
Kuaizhou-1                              & 0                                               & 621                                             & 686                                                 & 183                                                &                                                 & \cellcolor[HTML]{FFEB9C}{\color[HTML]{9C6500} \textless{}3.8}    & \cellcolor[HTML]{FFEB9C}{\color[HTML]{9C6500} \textgreater{}3.8} & orbital                                                          &          & 686                                                                                                                   \\
Kuaizhou-11                             & 0                                               & 7530                                            & 1490                                                & 520                                                &                                                 & \cellcolor[HTML]{FFEB9C}{\color[HTML]{9C6500} \textless{}3.8}    & \cellcolor[HTML]{FFEB9C}{\color[HTML]{9C6500} \textgreater{}3.8} & orbital                                                          &          & 1490                                                                                                                  \\
LauncherOne                             & 0                                               & 3000                                            & 250                                                 &                                                    &                                                 & 3.4                                                              & orbital                                                          &                                                                  &          & 0                                                                                                                     \\
Long March (CZ) 11                      & 0                                               & 5000                                            & \cellcolor[HTML]{FFC7CE}{\color[HTML]{9C0006} }     & \cellcolor[HTML]{FFEB9C}{\color[HTML]{9C6500} 500} & \cellcolor[HTML]{FFC7CE}{\color[HTML]{9C0006} } & \cellcolor[HTML]{FFEB9C}{\color[HTML]{9C6500} \textless{}3.8}    & \cellcolor[HTML]{FFEB9C}{\color[HTML]{9C6500} \textless{}3.8}    & \cellcolor[HTML]{FFEB9C}{\color[HTML]{9C6500} \textgreater{}3.8} & orbital  & 500                                                                                                                   \\
Long March (CZ) 12                      & 0                                               & 10000                                           & \cellcolor[HTML]{FFC7CE}{\color[HTML]{9C0006} }     &                                                    &                                                 & \cellcolor[HTML]{FFEB9C}{\color[HTML]{9C6500} \textless{}3.8}    & orbital                                                          &                                                                  &          & 0                                                                                                                     \\
Long March (CZ) 2C                      & 0                                               & 10150                                           & 5143                                                &                                                    &                                                 & 1.9                                                              & orbital                                                          &                                                                  &          & 0                                                                                                                     \\
Long March (CZ) 2C (SMA and YZ-1S)      & 0                                               & 10150                                           & 5143                                                & \cellcolor[HTML]{C6EFCE}{\color[HTML]{006100} }    &                                                 & 1.9                                                              & 7.9                                                              & orbital                                                          &          & 5143                                                                                                                  \\
Long March (CZ) 2D                      & 0                                               & 10443                                           & 4019                                                &                                                    &                                                 & 2.7                                                              & orbital                                                          &                                                                  &          & 0                                                                                                                     \\
Long March (CZ) 2D/YZ-3                 & 0                                               & 10443                                           & 4019                                                & \cellcolor[HTML]{FFC7CE}{\color[HTML]{9C0006} }    &                                                 & 2.7                                                              & 7.9                                                              & orbital                                                          &          & 4019                                                                                                                  \\
Long March (CZ) 2F                      & 13696                                           & 10165                                           & 8338                                                &                                                    &                                                 & 2.2                                                              & orbital                                                          &                                                                  &          & 0                                                                                                                     \\
Long March (CZ) 3A                      & 0                                               & 11021                                           & 5773                                                & 4743                                               &                                                 & \cellcolor[HTML]{FFEB9C}{\color[HTML]{9C6500} \textless{}3.8}    & \cellcolor[HTML]{FFEB9C}{\color[HTML]{9C6500} \textgreater{}3.8} & orbital                                                          &          & 5773                                                                                                                  \\
Long March (CZ) 3B and 3B/E             & 13696                                           & 11021                                           & 5773                                                & 4743                                               &                                                 & 2.6                                                              & 5.0                                                              & orbital                                                          &          & 5773                                                                                                                  \\
Long March (CZ) 3B/YZ-1                 & 13696                                           & 11021                                           & 5773                                                & 4743                                               & \cellcolor[HTML]{FFC7CE}{\color[HTML]{9C0006} } & 2.6                                                              & 5.0                                                              & 9.3                                                              & orbital  & 10515                                                                                                                 \\
Long March (CZ) 3C and 3C/E             & 6400                                            & 10300                                           & 5395                                                & 4432                                               &                                                 & 1.9                                                              & 4.2                                                              & orbital                                                          &          & 5395                                                                                                                  \\
Long March (CZ) 3C/YZ-1                 & 6400                                            & 10300                                           & 5395                                                & 4432                                               & \cellcolor[HTML]{FFC7CE}{\color[HTML]{9C0006} } & \cellcolor[HTML]{FFEB9C}{\color[HTML]{9C6500} \textless{}3.8}    & \cellcolor[HTML]{FFEB9C}{\color[HTML]{9C6500} \textless{}3.8}    & \cellcolor[HTML]{FFEB9C}{\color[HTML]{9C6500} \textgreater{}3.8} & orbital  & 4432                                                                                                                  \\
Long March (CZ) 4B                      & 0                                               & 10632                                           & 3159                                                & 3460                                               & 1488                                            & 2.1                                                              & 4.6                                                              & orbital                                                          & 4.2 (PF) & 4646                                                                                                                  \\
Long March (CZ) 4C                      & 0                                               & 10290                                           & 3041                                                & 3331                                               & 1488                                            & 2.1                                                              & 4.7                                                              & orbital                                                          & 4.3 (PF) & 4528                                                                                                                  \\
Long March (CZ) 5                       & 50310                                           & 19135                                           & 10665                                               &                                                    &                                                 & 6.3                                                              & orbital                                                          &                                                                  &          & 19135                                                                                                                 \\
Long March (CZ) 5/YZ-2                  & 50310                                           & 19135                                           & 10665                                               & \cellcolor[HTML]{FFC7CE}{\color[HTML]{9C0006} }    &                                                 & 6.3                                                              & 10.1                                                             & orbital                                                          &          & 29800                                                                                                                 \\
Long March (CZ) 5B                      & 51480                                           & 25612                                           &                                                     &                                                    &                                                 & orbital                                                          &                                                                  &                                                                  &          & 0    \\ \hline                                                                                                                 
\end{tabular}
\end{table*}

\setcounter{table}{6}
\begin{table*}[!t]
\caption{\textit{continued.}}
\footnotesize
\begin{tabular}{l|l|llll|llll|l}
\hline
Rocket                                  &  \multicolumn{5}{c}{Stage mass (kg)}                                                                                                                                                                               & \multicolumn{4}{c}{\begin{tabular}[c]{@{}c@{}}Stage re-entry velocity \\ at 100\,km (km/s)\end{tabular}}                                                                                                       & \multicolumn{1}{c}{\begin{tabular}[c]{@{}c@{}}Total re-entry mass \\ with v\textgreater{}3.8\,km/s (kg)\end{tabular}} \\
Stage \#                 & 0                                                  & 1                                                    & 2                                                   & 3                                                  & 4                                               & 1                                                             & 2                                                                & 3                                                                & 4        &                                                                                                                       \\ \hline
Long March (CZ) 6                       & 0                                               & 8000                                                & 3000                                                & 1000                                               &      & \cellcolor[HTML]{FFEB9C}{\color[HTML]{9C6500} \textless{}3.8}    & \cellcolor[HTML]{FFEB9C}{\color[HTML]{9C6500} \textgreater{}3.8} & orbital                                                          &         & 3000                                                                                                                  \\
Long March (CZ) 6A                      & 4800                                            & 8000                                                & 3000                                                & 1000                                               &      & \cellcolor[HTML]{FFEB9C}{\color[HTML]{9C6500} \textless{}3.8}    & \cellcolor[HTML]{FFEB9C}{\color[HTML]{9C6500} \textgreater{}3.8} & orbital                                                          &         & 3000                                                                                                                  \\
Long March (CZ) 6C                      & 0                                               & 8000                                                & \cellcolor[HTML]{FFC7CE}{\color[HTML]{9C0006} }     &                                                    &      & \cellcolor[HTML]{FFEB9C}{\color[HTML]{9C6500} \textless{}3.8}    & orbital                                                          &                                                                  &         & 0                                                                                                                     \\
Long March (CZ) 7                       & 24000                                           & 12000                                               & \cellcolor[HTML]{FFC7CE}{\color[HTML]{9C0006} }     &                                                    &      & \cellcolor[HTML]{FFEB9C}{\color[HTML]{9C6500} \textgreater{}3.8} & orbital                                                          &                                                                  &         & 12000                                                                                                                 \\
Long March (CZ) 7/YZ-1A                 & 24000                                           & 12000                                               & 6000                                                & \cellcolor[HTML]{FFC7CE}{\color[HTML]{9C0006} }    &      & \cellcolor[HTML]{FFEB9C}{\color[HTML]{9C6500} \textless{}3.8}    & \cellcolor[HTML]{FFEB9C}{\color[HTML]{9C6500} \textgreater{}3.8} & orbital                                                          &         & 6000                                                                                                                  \\
Long March (CZ) 7A                      & 24000                                           & 12000                                               & 6000                                                & 2800                                               &      & \cellcolor[HTML]{FFEB9C}{\color[HTML]{9C6500} \textless{}3.8}    & \cellcolor[HTML]{FFEB9C}{\color[HTML]{9C6500} \textgreater{}3.8} & orbital                                                          &         & 6000                                                                                                                  \\
Long March (CZ) 8 and 8A                & 12000                                           & 12000                                               & 2800                                                &                                                    &      & \cellcolor[HTML]{FFEB9C}{\color[HTML]{9C6500} \textgreater{}3.8} & orbital                                                          &                                                                  &         & 12000                                                                                                                 \\
Minotaur 1                              & 0                                               & 2292                                                & 795                                                 & 404                                                & 153  & \cellcolor[HTML]{FFEB9C}{\color[HTML]{9C6500} \textless{}3.8}    & \cellcolor[HTML]{FFEB9C}{\color[HTML]{9C6500} \textless{}3.8}    & \cellcolor[HTML]{FFEB9C}{\color[HTML]{9C6500} \textgreater{}3.8} & orbital & 404                                                                                                                   \\
Minotaur-4                              & 0                                               & 3610                                                & 3190                                                & 645                                                & 256  & \cellcolor[HTML]{FFEB9C}{\color[HTML]{9C6500} \textless{}3.8}    & \cellcolor[HTML]{FFEB9C}{\color[HTML]{9C6500} \textless{}3.8}    & \cellcolor[HTML]{FFEB9C}{\color[HTML]{9C6500} \textgreater{}3.8} & orbital & 645                                                                                                                   \\
Minotaur-4 Orion-38                     & 0                                               & 3610                                                & 3190                                                & 645                                                & 410  & \cellcolor[HTML]{FFEB9C}{\color[HTML]{9C6500} \textless{}3.8}    & \cellcolor[HTML]{FFEB9C}{\color[HTML]{9C6500} \textless{}3.8}    & \cellcolor[HTML]{FFEB9C}{\color[HTML]{9C6500} \textgreater{}3.8} & orbital & 645                                                                                                                   \\
NK Kerolox LV                           & 0                                               & \cellcolor[HTML]{FFC7CE}{\color[HTML]{9C0006} }     & \cellcolor[HTML]{FFEB9C}{\color[HTML]{9C6500} 1000} & \cellcolor[HTML]{FFC7CE}{\color[HTML]{9C0006} }    &      & \cellcolor[HTML]{FFEB9C}{\color[HTML]{9C6500} \textless{}3.8}    & \cellcolor[HTML]{FFEB9C}{\color[HTML]{9C6500} \textgreater{}3.8} & orbital                                                          &         & 1000                                                                                                                  \\
New Glenn                               & 0                                               & 110000                                              & \cellcolor[HTML]{FFC7CE}{\color[HTML]{9C0006} }     &                                                    &      & lands                                                            & orbital                                                          &                                                                  &         & 0                                                                                                                     \\
Nuri                                    & 0                                               & 16000                                               & 4400                                                & 1290                                               &      & \cellcolor[HTML]{FFEB9C}{\color[HTML]{9C6500} \textless{}3.8}    & \cellcolor[HTML]{FFEB9C}{\color[HTML]{9C6500} \textgreater{}3.8} & orbital                                                          &         & 4400                                                                                                                  \\
PSLV (all types)                        & \cellcolor[HTML]{C6EFCE}{\color[HTML]{006100} } & 30100                                               & 5350                                                & 900                                                & 670  & \textless{}3.8                                                   & 4.0                                                              & 6.1                                                              & orbital & 6250                                                                                                                  \\
Pegasus XL                              & 0                                               & 1386                                                & 416                                                 & 108                                                &      & 2.3                                                              & 5.6                                                              & orbital                                                          &         & 416                                                                                                                   \\
Proton-M                                & 0                                               & 30600                                               & 11000                                               & 3500                                               &      & \textless{}3.8                                                   & 4.6                                                              & orbital                                                          &         & 11000                                                                                                                 \\
Proton-M (Briz-M and DM-3)              & 0                                               & 30600                                               & 11000                                               & 3500                                               & 2370 & \textless{}3.8                                                   & 4.6                                                              & 7.3                                                              & orbital & 14500                                                                                                                 \\
Qaem-100                                & 0                                               & \cellcolor[HTML]{FFEB9C}{\color[HTML]{9C6500} 3030} & \cellcolor[HTML]{FFEB9C}{\color[HTML]{9C6500} 465}  & \cellcolor[HTML]{FFC7CE}{\color[HTML]{9C0006} }    &      & \cellcolor[HTML]{FFEB9C}{\color[HTML]{9C6500} \textless{}3.8}    & \cellcolor[HTML]{FFEB9C}{\color[HTML]{9C6500} \textgreater{}3.8} & orbital                                                          &         & 465                                                                                                                   \\
Qased                                   & 0                                               & 2700                                                & 200                                                 & 133                                                &      & \cellcolor[HTML]{FFEB9C}{\color[HTML]{9C6500} \textless{}3.8}    & \cellcolor[HTML]{FFEB9C}{\color[HTML]{9C6500} \textgreater{}3.8} & orbital                                                          &         & 200                                                                                                                   \\
Rokot-KM                                & 0                                               & \cellcolor[HTML]{FFC7CE}{\color[HTML]{9C0006} }     & 1500                                                & \cellcolor[HTML]{FFC7CE}{\color[HTML]{9C0006} }    &      & \textless{}3.8                                                   & 5.9                                                              & orbital                                                          &         & 1500                                                                                                                  \\
SS-520                                  & 0                                               & 400                                                 & 111                                                 & 8                                                  &      & \cellcolor[HTML]{FFEB9C}{\color[HTML]{9C6500} \textless{}3.8}    & \cellcolor[HTML]{FFEB9C}{\color[HTML]{9C6500} \textgreater{}3.8} & orbital                                                          &         & 111                                                                                                                   \\
SSLV                                    & 0                                               & 10500                                               & 2650                                                & 1450                                               & 150  & 2.1                                                              & 4.1                                                              & 7.6                                                              & orbital & 4100                                                                                                                  \\
Safir                                   & 0                                               & 3030                                                & 465                                                 &                                                    &      & \cellcolor[HTML]{FFEB9C}{\color[HTML]{9C6500} \textless{}3.8}    & orbital                                                          &                                                                  &         & 0                                                                                                                     \\
Shavit and Shavit 2                     & 0                                               & 1240                                                & 1376                                                & 684                                                &      & \cellcolor[HTML]{FFEB9C}{\color[HTML]{9C6500} \textless{}3.8}    & \cellcolor[HTML]{FFEB9C}{\color[HTML]{9C6500} \textgreater{}3.8} & orbital                                                          &         & 1376                                                                                                                  \\
Shuang Quxian-1 and 1A                  & 0                                               & 1744                                                & 974                                                 & 282                                                & 300  & \cellcolor[HTML]{FFEB9C}{\color[HTML]{9C6500} \textless{}3.8}    & \cellcolor[HTML]{FFEB9C}{\color[HTML]{9C6500} \textless{}3.8}    & \cellcolor[HTML]{FFEB9C}{\color[HTML]{9C6500} \textgreater{}3.8} & orbital & 282                                                                                                                   \\
Simorgh                                 & 0                                               & 12700                                               & 1600                                                & 230                                                &      & \cellcolor[HTML]{FFEB9C}{\color[HTML]{9C6500} \textless{}3.8}    & \cellcolor[HTML]{FFEB9C}{\color[HTML]{9C6500} \textgreater{}3.8} & orbital                                                          &         & 1600                                                                                                                  \\
Soyuz-2-1A                              & 15136                                           & 6545                                                & 2410                                                &                                                    &      & 4.3                                                              & orbital                                                          &                                                                  &         & 6545                                                                                                                  \\
Soyuz-2-1A (Fregat, Fregat-M and Volga) & 15136                                           & 6545                                                & 2410                                                & \cellcolor[HTML]{C6EFCE}{\color[HTML]{006100} }    &      & 3.9                                                              & 7.0                                                              & orbital                                                          &         & 8955                                                                                                                  \\
Soyuz-2-1B                              & 15136                                           & 6545                                                & 2710                                                &                                                    &      & 4.3                                                              & orbital                                                          &                                                                  &         & 6545                                                                                                                  \\
Soyuz-2-1B (Fregat and Fregat-M)        & 15136                                           & 6545                                                & 2355                                                & \cellcolor[HTML]{C6EFCE}{\color[HTML]{006100} }    &      & 3.9                                                              & 7.0                                                              & orbital                                                          &         & 8900                                                                                                                  \\
Soyuz-2-1V                              & 0                                               & 9300                                                & 2355                                                &                                                    &      & \cellcolor[HTML]{FFEB9C}{\color[HTML]{9C6500} \textless{}3.8}    & orbital                                                          &                                                                  &         & 0                                                                                                                     \\
Soyuz-2-1V Volga                        & 15136                                           & 9300                                                & 2355                                                & 890                                                &      & \cellcolor[HTML]{FFEB9C}{\color[HTML]{9C6500} \textless{}3.8}    & \cellcolor[HTML]{FFEB9C}{\color[HTML]{9C6500} \textgreater{}3.8} & orbital                                                          &         & 2355                                                                                                                  \\
Soyuz-FG                                & 15136                                           & 6545                                                & \cellcolor[HTML]{C6EFCE}{\color[HTML]{006100} }     &                                                    &      & 3.9                                                              & orbital                                                          &                                                                  &         & 6545                                                                                                                  \\
Soyuz-ST-A Fregat-M and ST-B Fregat-MT  & 15136                                           & 6545                                                & 2470                                                & \cellcolor[HTML]{C6EFCE}{\color[HTML]{006100} }    &      & 3.9                                                              & 7.0                                                              & orbital                                                          &         & 9015                                                                                                                  \\
Soyuz-U and Soyuz-U PVB                 & 15136                                           & 6545                                                & 2410                                                &                                                    &      & 4.3                                                              & orbital                                                          &                                                                  &         & 6545                                                                                                                  \\
Space Launch System - Block 1 Crew      & 198600                                          & 99300                                               & 3700                                                &                                                    &      & 8.3                                                              & orbital                                                          &                                                                  &         & 99300                                                                                                                 \\
Starship                                & 0                                               & 220000                                              & 120000                                              &                                                    &      & lands                                                            & orbital                                                          &                                                                  &         & 0                                                                                                                     \\
Starship V2                             & 0                                               & 220000                                              & 127200                                              &                                                    &      & lands                                                            & orbital                                                          &                                                                  &         & 0                                                                                                                     \\
Taurus                                  & 0                                               & 4450                                                & 1700                                                & 420                                                & 126  & 1.6                                                              & 3.9                                                              & 7.1                                                              & orbital & 2120                                                                                                                  \\
Tianlong 2                              & 0                                               & \cellcolor[HTML]{FFEB9C}{\color[HTML]{9C6500} 9300} & \cellcolor[HTML]{FFEB9C}{\color[HTML]{9C6500} 2355} & \cellcolor[HTML]{FFEB9C}{\color[HTML]{9C6500} 890} &      & \cellcolor[HTML]{FFEB9C}{\color[HTML]{9C6500} \textless{}3.8}    & \cellcolor[HTML]{FFEB9C}{\color[HTML]{9C6500} \textgreater{}3.8} & orbital                                                          &         & 2355                                                                                                                  \\
Unha-3                                  & 0                                               & 8000                                                & 1000                                                & 900                                                &      & 2.8                                                              & 5.4                                                              & orbital                                                          &         & 1000                                                                                                                  \\
Vega                                    & 0                                               & 8533                                                & 2486                                                & 1433                                               & 688  & \textless{}3.8                                                   & 4.0                                                              & 7.6                                                              & orbital & 3919                                                                                                                  \\
Vega C                                  & 0                                               & 13393                                               & 4238                                                & 1433                                               & 698  & \textless{}3.8                                                   & 4.6                                                              & 7.6                                                              & orbital & 5671                                                                                                                  \\
Vulcan Centaur VC2S                     & 10400                                           & 50000                                               & 5500                                                &                                                    &      & 3.3                                                              & orbital                                                          &                                                                  &         & 0                                                                                                                     \\
Zenit-3F                                & 0                                               & 33900                                               & 9300                                                & 1050                                               &      & \cellcolor[HTML]{FFEB9C}{\color[HTML]{9C6500} \textless{}3.8}    & \cellcolor[HTML]{FFEB9C}{\color[HTML]{9C6500} \textgreater{}3.8} & orbital                                                          &         & 9300                                                                                                                  \\
Zhongke 1A                              & 0                                               & 13393                                               & 4238                                                & 1433                                               & 698  & \cellcolor[HTML]{FFEB9C}{\color[HTML]{9C6500} \textless{}3.8}    & \cellcolor[HTML]{FFEB9C}{\color[HTML]{9C6500} \textgreater{}3.8} & \cellcolor[HTML]{FFEB9C}{\color[HTML]{9C6500} \textgreater{}3.8} & orbital & 5671                                                                                                                  \\
Zhuque-1                                & 0                                               & \cellcolor[HTML]{FFC7CE}{\color[HTML]{9C0006} }     & \cellcolor[HTML]{FFEB9C}{\color[HTML]{9C6500} 500}  & \cellcolor[HTML]{FFC7CE}{\color[HTML]{9C0006} }    &      & \cellcolor[HTML]{FFEB9C}{\color[HTML]{9C6500} \textless{}3.8}    & \cellcolor[HTML]{FFEB9C}{\color[HTML]{9C6500} \textgreater{}3.8} & orbital                                                          &         & 500                                                                                                                   \\
Zhuque-2                                & 0                                               & 19700                                               & 1392                                                &                                                    &      & \cellcolor[HTML]{FFEB9C}{\color[HTML]{9C6500} \textless{}3.8}    & orbital                                                          &                                                                  &         & 0                                                                                                                    \\ \hline                                                                                                               
\end{tabular}
\end{table*}

\section*{Appendix B: Satellite and rocket body average elemental composition} \label{AppB}
Table \ref{tab:AppB} gives the average elemental mass composition estimates for different types of space waste, namely satellites, the 3 different rocket upper stages, and rocket core stages. The compositions are calculated by combining the material group/component mass fraction estimates with the elemental mass composition of the respective materials and components (both are provided in Tables \ref{tab:sat_comp} for satellites and \ref{tab:RB_comp} for the 4 different rocket body types). The color code in Table \ref{tab:AppB} describes the level of uncertainty introduced in Section \ref{sec:errors}.

\begin{table*}[!h]
\centering
\caption{Average elemental mass composition estimates of the different types of space waste. Elements marked green reflect the uncertainty levels described in Section \ref{sec:errors}; green implies a $3\sigma$-uncertainty equal to 40\% of the elemental mass fraction, yellow 80\%. Elements with very high uncertainty are summarized under ''Others``. Values are given in \%.}
\label{tab:AppB}
\begin{tabular}{%
    l  
    d d d d d      
}
\hline
Element
  & \multicolumn{1}{c}{Satellite}
  & \multicolumn{1}{c}{\begin{tabular}[c]{@{}c@{}}Average\\upper stage\end{tabular}}
  & \multicolumn{1}{c}{\begin{tabular}[c]{@{}c@{}}Falcon 9/Heavy\\upper stage\end{tabular}}
  & \multicolumn{1}{c}{\begin{tabular}[c]{@{}c@{}}Starship\\upper stage\end{tabular}}
  & \multicolumn{1}{c}{Core stage} \\\hline
\cellcolor[HTML]{C6EFCE}{\color[HTML]{006100} H}  & 1.50       & 0.88                & 0.72                       & 0.22                 & 0.61       \\
\cellcolor[HTML]{C6EFCE}{\color[HTML]{006100} Li} & 0.27       & 0.08                & 0.35                       & 0.02                 & 0.08       \\
\cellcolor[HTML]{FFEB9C}{\color[HTML]{9C6500} Be} & 0.01       & 0.00                & 0.00                       & 0.00                 & 0.00       \\
\cellcolor[HTML]{C6EFCE}{\color[HTML]{006100} B}  & 0.10       & 0.03                & 0.03                       & 0.02                 & 0.02       \\
\cellcolor[HTML]{C6EFCE}{\color[HTML]{006100} C}  & 24.11      & 15.07               & 10.01                      & 4.43                 & 11.22      \\
\cellcolor[HTML]{C6EFCE}{\color[HTML]{006100} N}  & 0.90       & 0.83                & 0.68                       & 0.34                 & 0.72       \\
\cellcolor[HTML]{C6EFCE}{\color[HTML]{006100} O}  & 8.87       & 4.40                & 3.56                       & 4.81                 & 3.04       \\
\cellcolor[HTML]{FFEB9C}{\color[HTML]{9C6500} F}  & 2.36       & 0.41                & 0.39                       & 0.23                 & 0.28       \\
\cellcolor[HTML]{FFEB9C}{\color[HTML]{9C6500} Na} & 0.06       & 0.00                & 0.00                       & 0.00                 & 0.00       \\
\cellcolor[HTML]{C6EFCE}{\color[HTML]{006100} Mg} & 1.18       & 0.36                & 0.39                       & 0.06                 & 0.28       \\
\cellcolor[HTML]{C6EFCE}{\color[HTML]{006100} Al} & 35.99      & 49.63               & 57.78                      & 5.32                 & 50.69      \\
\cellcolor[HTML]{C6EFCE}{\color[HTML]{006100} Si} & 2.81       & 1.61                & 1.49                       & 3.99                 & 1.41       \\
\cellcolor[HTML]{C6EFCE}{\color[HTML]{006100} P}  & 0.34       & 0.04                & 0.04                       & 0.04                 & 0.03       \\
\cellcolor[HTML]{FFEB9C}{\color[HTML]{9C6500} S}  & 0.02       & 0.01                & 0.01                       & 0.01                 & 0.01       \\
\cellcolor[HTML]{FFEB9C}{\color[HTML]{9C6500} Cl} & 0.06       & 0.04                & 0.03                       & 0.01                 & 0.03       \\
\cellcolor[HTML]{FFEB9C}{\color[HTML]{9C6500} K}  & 0.06       & 0.00                & 0.00                       & 0.00                 & 0.00       \\
\cellcolor[HTML]{FFEB9C}{\color[HTML]{9C6500} Ca} & 0.19       & 0.10                & 0.10                       & 0.06                 & 0.07       \\
\cellcolor[HTML]{C6EFCE}{\color[HTML]{006100} Ti} & 4.33       & 4.34                & 4.11                       & 3.11                 & 3.62       \\
\cellcolor[HTML]{FFEB9C}{\color[HTML]{9C6500} V}  & 0.15       & 0.21                & 0.18                       & 0.13                 & 0.18       \\
\cellcolor[HTML]{C6EFCE}{\color[HTML]{006100} Cr} & 0.60       & 1.90                & 0.99                       & 13.15                & 2.69       \\
\cellcolor[HTML]{C6EFCE}{\color[HTML]{006100} Mn} & 0.10       & 0.23                & 0.23                       & 0.66                 & 0.26       \\
\cellcolor[HTML]{C6EFCE}{\color[HTML]{006100} Fe} & 2.92       & 7.48                & 4.32                       & 46.28                & 9.91       \\
\cellcolor[HTML]{FFEB9C}{\color[HTML]{9C6500} Co} & 0.83       & 0.66                & 1.04                       & 0.68                 & 0.74       \\
\cellcolor[HTML]{C6EFCE}{\color[HTML]{006100} Ni} & 1.30       & 2.60                & 2.73                       & 10.17                & 4.15       \\
\cellcolor[HTML]{C6EFCE}{\color[HTML]{006100} Cu} & 6.87       & 5.99                & 4.81                       & 4.03                 & 6.82       \\
\cellcolor[HTML]{C6EFCE}{\color[HTML]{006100} Zn} & 1.00       & 1.05                & 0.79                       & 0.05                 & 0.75       \\
\cellcolor[HTML]{FFEB9C}{\color[HTML]{9C6500} Ga} & 0.01       & 0.00                & 0.00                       & 0.00                 & 0.00       \\
\cellcolor[HTML]{FFEB9C}{\color[HTML]{9C6500} Ge} & 0.48       & 0.00                & 0.00                       & 0.00                 & 0.00       \\
\cellcolor[HTML]{FFEB9C}{\color[HTML]{9C6500} As} & 0.01       & 0.00                & 0.00                       & 0.00                 & 0.00       \\
\cellcolor[HTML]{FFEB9C}{\color[HTML]{9C6500} Br} & 0.19       & 0.12                & 0.12                       & 0.07                 & 0.08       \\
\cellcolor[HTML]{FFEB9C}{\color[HTML]{9C6500} Zr} & 0.03       & 0.05                & 0.08                       & 0.01                 & 0.06       \\
\cellcolor[HTML]{FFEB9C}{\color[HTML]{9C6500} Nb} & 0.00       & 0.57                & 3.33                       & 1.06                 & 1.06       \\
\cellcolor[HTML]{C6EFCE}{\color[HTML]{006100} Mo} & 0.03       & 0.11                & 0.10                       & 0.20                 & 0.20       \\
\cellcolor[HTML]{FFEB9C}{\color[HTML]{9C6500} Ag} & 0.81       & 0.07                & 0.14                       & 0.04                 & 0.07       \\
\cellcolor[HTML]{FFEB9C}{\color[HTML]{9C6500} ln} & 0.00       & 0.00                & 0.00                       & 0.00                 & 0.00       \\
\cellcolor[HTML]{C6EFCE}{\color[HTML]{006100} Sn} & 0.86       & 0.54                & 0.52                       & 0.31                 & 0.38       \\
\cellcolor[HTML]{FFEB9C}{\color[HTML]{9C6500} Ba} & 0.08       & 0.02                & 0.02                       & 0.01                 & 0.01       \\
\cellcolor[HTML]{FFEB9C}{\color[HTML]{9C6500} La} & 0.00       & 0.00                & 0.00                       & 0.00                 & 0.00       \\
\cellcolor[HTML]{FFEB9C}{\color[HTML]{9C6500} Ce} & 0.01       & 0.00                & 0.00                       & 0.00                 & 0.00       \\
\cellcolor[HTML]{FFEB9C}{\color[HTML]{9C6500} Hf} & 0.00       & 0.05                & 0.36                       & 0.10                 & 0.10       \\
\cellcolor[HTML]{FFEB9C}{\color[HTML]{9C6500} Ta} & 0.01       & 0.01                & 0.03                       & 0.01                 & 0.01       \\
\cellcolor[HTML]{FFEB9C}{\color[HTML]{9C6500} W}  & 0.00       & 0.15                & 0.17                       & 0.15                 & 0.16       \\
\cellcolor[HTML]{FFEB9C}{\color[HTML]{9C6500} Pb} & 0.53       & 0.36                & 0.35                       & 0.20                 & 0.25       \\
Others                                            & 0.02       & 0.01                & 0.01                       & 0.00                 & 0.01       \\ \hline
Sum                                               & 100.00     & 100.00              & 100.00                     & 100.00               & 100.00    \\ \hline
\end{tabular}
\end{table*}


\bibliographystyle{jasr-model5-names}
\biboptions{authoryear}
\bibliography{bib}

\end{document}